\documentclass[journal]{IEEEtran}

\usepackage{times,graphicx,epsf,epsfig,color,cite,psfrag}
\usepackage{amsfonts,amssymb,amsmath,amssymb}
\usepackage{algorithm}
\usepackage[noend]{algpseudocode}

\def\reals{ { {\rm  I \kern-0.15em R }  } }
\def\complex{ {\,{{\rm C} \kern-0.50em \raise0.20ex {  |}}\, }}

\def\hbf{{\bf h}}

\def\sbf{{\bf s}}

\def\Hbf{{\bf H}}

\def\Ac{{\cal A}}

\def\Oc{{\cal O}}

\def\eg{{\it e.g.,\ \/}}

\def\ie{{\it i.e.,\ \/}}

\newcommand{\mbbE}{\mathbb{E}}
\def\nn{\nonumber}

\def\topt{\text{opt}}

\newtheorem{theorem}{Theorem}
\newtheorem{corollary}{Corollary}
\newtheorem{lemma}{Lemma}

\newtheorem{proposition}{Proposition}

\usepackage[normalem]{ulem}

\begin{document}

\title{ Online Power Control Optimization for Wireless Transmission with Energy Harvesting and Storage}
\author{Fatemeh Amirnavaei, \IEEEmembership{Student Member, IEEE} and Min Dong\thanks{This work was supported in part by the Ontario Ministry of Research and Innovation under an Early Researcher Award, and in part by the National Sciences and Engineering Research Council (NSERC) of Canada under Discovery Grant RGPIN-2014-05181. A preliminary version of this work \cite{AmirDong:SPAWC15} was presented at \emph{the 2015 IEEE International Workshop on Signal Processing Advances in Wireless Communications} (SPAWC), Stockholm,  June 2015.}\thanks{The authors are with Department of Electrical, Computer and Software Engineering, University of Ontario Institute of Technology, Ontario, Canada (email: \{fatemeh.amirnavaei,~min.dong\}@uoit.ca).}, \IEEEmembership{Senior Member, IEEE}}
\maketitle
\begin{abstract}
We consider wireless transmission over fading channel powered by energy harvesting and storage devices. Assuming a finite battery storage capacity, we design an online power control strategy aiming at maximizing the long-term time-averaged transmission rate under battery operational constraints for energy harvesting.  We first formulate the stochastic optimization problem, and then develop techniques to transform this problem and employ techniques from Lyapunov optimization to design the online power control solution.
In particular, we propose an approach to handle unbounded channel   fade which cannot by directly dealt with by Lyapunov framework. Our proposed algorithm determines the transmission power based only on the current energy state of the battery and channel fade conditions,without requiring any knowledge of the statistics of energy arrivals and fading channels. Our online power control solution is a three-stage  closed-form solution depending on the battery energy level. It not only  provides strategic energy conservation through the battery energy control, but also reveals  an opportunistic transmission  style based on fading
condition, both of which improve the long-term time-averaged transmission rate.
We further characterize the performance bound of our proposed algorithm to the optimal solution with a general fading distribution. Simulation results demonstrate a significant performance gain of our proposed online algorithm over alternative online approaches.
\end{abstract}

\section{Introduction} \label{sec:intro}
The excess carbon emission due to the growing energy demand has caused significant environmental concern.  To face increasing energy cost and reduce carbon footprint, renewable generation has increasingly been considered as an alternative energy source for power supply in wireless communication systems. In particular,
using energy harvesting devices to supply power to wireless transmitters  has recently attracted a growing attention. Unlike the traditional fixed power supply either from the grid or battery, an energy harvesting device can scavenge energy from renewable energy sources in the environment and provide continuous power supplies. There is a growing demand of this low-cost green technology for a wide range of applications, such as supplying power to  base station or relay station in cellular networks. For energy-constrained wireless applications such as sensor networks, energy harvesting device provides an unlimited power supply to maintain the lifetime of the network operation without the need to replenish batteries.

For wireless transmission powered by renewable energy, typically, an energy harvesting device is implemented with a storage   battery to store the harvested energy and to provide power  for transmission. With energy harvesting and storage, power control for  transmission over  fading channels faces unique challenges, including the randomness of  both renewable energy source and wireless fading channels, and the battery operational constraints on energy harvesting and power supply.

To address this problem,  several existing works have considered off-line optimal power control designs  for   additive white Gaussian noise (AWGN) channels \cite{OzelUlukus:JIT12,YangUlukus:JCOM12,HuangCui13,TutuncuogluYener:TWCOM12,TutunYener12} and for fading channels \cite{HoZhang:TSP12,Yuanetal:TWCOM15,OzelYener:TSA11}, where the  harvested energy  (and channel fades in case of fading channels) within a time period are assumed known beforehand.     However, for practical system designs, both harvested energy and channel quality can only be acquired causally. Some existing works have proposed online power control strategies \cite{WangTajer:JCOM12,HoZhang:TSP12,OzelYener:TSA11,Sharmaetal:TWCOM10,Jing:JWCOM09,BlascoGunduz:JWCOM13,TutYener:CISS12,Zhang:JSAC14,OrGunduz:ICC13, OzelTut:JSAC11} based on the current and past system information. However, these works also assume statistics of energy arrivals and channel fades to be certain types and known at the transmitter, and the solutions often need to be obtained numerically with high computational complexity. In reality, the statistics of the energy arrival for harvesting are  difficult to obtain or predict accurately.    Thus, it is desirable and practical to design online power control which only relies on the harvested energy and fading condition up to the current time without requiring their statistical knowledge. In addition, most existing works often make simple assumptions on the battery operation for energy harvesting and power supply, without realistically consider battery operational constraints. However, these constraints limit the amount of energy that can be stored or drawn, and affect the transmission performance. Thus, a more realistic battery operation model for energy harvesting and power supply should be considered in the  power control design.

\subsection{Contributions}\label{subsec:intro.A}
In this paper, we consider the problem of power control for transmission powered by energy harvesting and storage devices for transmission over fading channels. For energy harvesting, we consider  a finite battery storage capacity and model the battery operational constraints on charging and power output.
In addition, we assume the statistics of energy arrivals and fading are unknown at the transmitter. Our goal is to  maximize the long-term time-averaged transmission rate under the battery operational constraints.

Our formulated optimization problem is stochastic and technically challenging to solve. In particular, the finite battery storage capacity and operational constraints cause the power control decision coupled over time which complicates the control decision making. We leverage  Lyapunov optimization framework \cite{Neely10} to design online power control. However,  applying the Lyapunov  technique to our problem is nontrivial. Specifically,  the original optimization problem cannot be directly handled by Lyapunov framework. Several issues need to be addressed, including the form of  battery operational constraints and unbounded channel fades. We develop special techniques to handle these issues to tackle the online power control problem. Our main contributions are summarized as follows:
\begin{itemize}
\item We formulate the transmission power control under energy harvesting and storage for the long-term average rate maximization  over fading channels as a stochastic optimization problem by taking into account detailed battery operational dynamics and constraints.\item We propose, to the best of our knowledge, the first online power control algorithm under realistic battery operational dynamics and constraints for transmission over fading channels. Our proposed algorithm determines transmit power  only based on the current energy state of battery and fading condition, without requiring any statistical knowledge of energy arrivals and channel fades. Our online solution is given in closed-form which is not only simple to implement, but also provides insight of the energy conservation and control in the battery and transmission control over fading.
In particular, transmission under our power control solution  turns out to be in  an opportunistic fashion based on the fading condition and the battery energy level, resembling a ``water-filling" like solution.
Furthermore, although focusing on a single-antenna transmission  system, we show that our proposed online power control algorithm is applicable to general  multi-antenna beamforming  scenarios.

\item We analyze our proposed algorithm and  show that it  has a bounded performance gap to the optimal solution with a general fading distribution.
\item We study the performance of our proposed online power control algorithm via simulation and demonstrate that a significant gain is achieved by our proposed algorithm over several alternative algorithms. We further numerically analyze our proposed algorithm  under different battery storage size, energy arrival rate, and fading conditions. In particular, we show that    it is near optimal even with relatively small battery storage size.  \end{itemize}

\subsection{Related Work}

Due to the randomness of the energy source and the wireless fading channels, existing works on the transmission power control design can be grouped into two categories: off-line and online power control strategies.
For an off-line power control design, energy arrivals and channel fades within a time period are known non-causally. In this case,  typically a deterministic power optimization problem can be formulated with various criteria.  Several  literature works have considered off-line strategies for AWGN channels  \cite{OzelUlukus:JIT12,YangUlukus:JCOM12,HuangCui13,TutuncuogluYener:TWCOM12,TutunYener12,Yuanetal:TWCOM15,HoZhang:TSP12,OzelYener:TSA11}. From information theoretic point of view, the capacity of the AWGN channel with an energy harvesting transmitter has been derived in \cite{OzelUlukus:JIT12}. Optimal power allocation solution to minimize the transmission time for point-to-point transmission has been obtained in \cite{YangUlukus:JCOM12}. Power allocation for throughput maximization in a Gaussian relay channel under energy harvesting has been considered in \cite{HuangCui13}.
In all these works,  infinite capacity of the battery is assumed for energy storage. With finite battery capacity, power allocation policy  for rate maximization has been investigated for both single user and two-user Gaussian interference channel \cite{TutuncuogluYener:TWCOM12,TutunYener12}. For fading channels, power allocation solution for throughput maximization has been obtained for infinite battery capacity \cite{HoZhang:TSP12,Yuanetal:TWCOM15} and finite battery capacity cases \cite{OzelYener:TSA11}. For \cite{Yuanetal:TWCOM15}, different from the commonly used harvest-store-use models for energy harvesting, the authors have considered a harvest-use-store model to improve the efficiency of energy usage.

Online power control design based on the current and past system information, such as energy arrivals, is a more practical but much challenging problem. A few existing works have formulated the power control problem by a Markov decision process (MDP) and obtain the power solution by Dynamic Programming (DP) for rate maximization or transmission error minimization \cite{WangTajer:JCOM12,HoZhang:TSP12,OzelYener:TSA11,Sharmaetal:TWCOM10,Jing:JWCOM09,BlascoGunduz:JWCOM13}.
For example, in \cite{HoZhang:TSP12}, online power control for rate maximization over a fading channel in finite time slots has been considered, where the harvested energy and fading are modeled as first-order Markov processes. To compute the power solution by DP, these works generally require the statistics of energy harvested and fading channel to be certain types and  known at the transmitter. In addition, the numerical solutions by DP are typically obtained with high computational complexity which is impractical for real implementation. Some low-complexity heuristic online approaches are proposed in \cite{OzelYener:TSA11,TutYener:CISS12,Zhang:JSAC14,OrGunduz:ICC13}. However, they also assume certain known statistical information and there is no performance guarantee. For the sensing application in a sensor network with energy harvesting, without the knowledge of energy arrival statistics, online power control to maximize long-term average sensing rate is considered for the AWGN\ channel  in \cite{MaoShroff:JAC12}, where Lyapunov technique is used in providing an online power solution. The maximization of utility performance for a network with   energy harvesting nodes is studied in  \cite{HuangNeely:JN13}, where an online algorithm based on Lyapunov technique is presented to jointly manage the energy and power allocation of packet transmissions.

Besides renewable sources such as solar and wind, harvesting energy from radio-frequency energy signals has recently been considered for wireless transfer of information and power simultaneously \cite{varshney:isit08,GoverSahi:isit10,Zhang:JWCOM13,XuZhang:TSP14,ShiZhang:JWCOM14,LeeZhang:JWCOM15}. First proposed in \cite{varshney:isit08}, simultaneous wireless information and power transfer (SWIPT) has been studied extensively under different system model assumptions. Point-to-point single antenna transmission is considered in \cite{varshney:isit08,GoverSahi:isit10}. A multiple input multiple output (MIMO) SWIPT system is first presented in~\cite{Zhang:JWCOM13}, and then is extended to a multiple input-single output (MISO) with more than two users in \cite{XuZhang:TSP14,ShiZhang:JWCOM14,LeeZhang:JWCOM15}.
No energy storage unit is considered in these works.

Given these recent works on energy harvesting, few studies have considered online power control over fading channels.  Different from most existing works for online power control design, we consider more sophisticated energy harvesting constraints due to battery charging and power output characteristics, and make no assumption on known prior statistics or distribution  of  energy harvested or fading. In addition, unlike those online solutions obtained by DP \cite{WangTajer:JCOM12,HoZhang:TSP12,OzelYener:TSA11,Sharmaetal:TWCOM10,Jing:JWCOM09,BlascoGunduz:JWCOM13} which suffer from high computational complexity, our online power solution is provided in closed-form  and thus very simple to implement.
Among the existing work, \cite{MaoShroff:JAC12} has used techniques related to Lyapunov framework to provide an online solution without requiring statistical knowledge of harvested energy. However,  the problem there is regarding  sensing rate maximization  by jointly controlling sensing rate and power allocation. The problem structure, formulation and constraints are very different from our work. Due to different form of constraints, the  approach and procedure to design rate control and power allocation through Lyapunov framework are quite different from ours.
Furthermore, \cite{MaoShroff:JAC12} only considers the AWGN\ channel case. As mentioned in Section~\ref{subsec:intro.A}, the consideration of the fading channel is highly nontrivial in both design and performance analysis, where Lyapunov optimization technique cannot be directly applied to this case with unbounded fading channel gain.

\subsection{Organization}   The rest of the paper is organized as follows. In Section \ref{sec:system model}, we provide the  system model. In Section \ref{sec:power control}, we formulate the power control optimization problem and propose our online power control algorithms for the point-to-point fading channel. Section \ref{sec:performance analysis} provides the performance analysis of our proposed algorithms. Section \ref{sec:simulation results} presents the simulation results, and Section \ref{sec:conclusion} concludes the paper.

\section{System Model}\label{sec:system model}
\begin{figure}[t]
\centerline{
\begin{psfrags}
\psfrag{Ea}[u]{$E_a(t)$}
\psfrag{gamma}[c]{\normalsize ${\gamma(t)}$}
\psfrag{Eb}[c]{\normalsize ${E_b(t)}$}
\psfrag{Eh}[c]{\normalsize ${E_s(t)}$}
\psfrag{Pt}[c]{ ${P(t)}$}
\psfrag{T}[c]{\normalsize T}
\psfrag{Energy}[l]{\footnotesize Energy}
\psfrag{Harvester}[l]{\footnotesize  Harvester}
\psfrag{R}[l]{\normalsize R}
\psfrag{Emax}[c]{\normalsize $E_{\max}$}
\psfrag{m}[c]{$\times$}
\psfrag{a}[c]{$+$}
\includegraphics[scale=0.6]{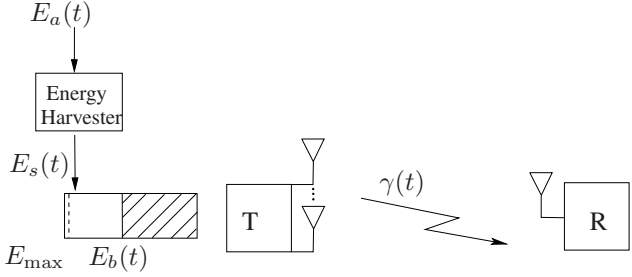}
\end{psfrags}
}
\caption{The  system model with energy harvesting and storage devices.}\label{fig:sysmod}
\end{figure}

We consider a point-to-point  wireless transmission system where the  transmitter is equipped with energy harvesting and storage devices as illustrated in Fig.~\ref{fig:sysmod}. The system operates in   discrete slotted time $t\in \{0,1,2,\ldots\}$ with duration $\Delta t$, and all operations are performed per time slot. The transmitter is powered by energy harvested from the environment (\eg solar, radio wave) using the harvesting device. Let $E_a(t)$  denote the amount of energy arrived at the harvesting device at time slot $t$, and $E_s(t)$ denote the amount of energy actually harvested into the battery at the end of time slot $t$. We have $E_s(t)\leq E_a(t)$. A battery storage device is used at the transmitter to store the harvested energy and to supply power for data transmission. Let $E_b(t)$ denote the energy state of battery (SOB) at the beginning of time slot $t$. It is bounded by
\begin{align}\label{batst}
E_{\min}\leq E_b(t)\leq E_{\max}, \ \forall t
\end{align}
where $E_{\min}$ and $E_{\max}$ represent the minimum and maximum energy levels allowed in the battery, respectively; their values depend on the type and size of the battery.

The battery has its maximum charging and discharging rates. Let $E_{c,\max}$ denote the maximum charging amount per slot. Let $P_{\max}$ denote the maximum transmit power that can be drawn from the battery, which should  satisfy
$\Delta tP_{\max} \le E_{\max}-E_{\min}$.
In addition, we assume $E_{c,\max}\le \Delta t P_{\max}$, \ie the maximum charging rate is no more than the maximum discharging rate\footnote{Based on the battery technology,  for current rechargeable batteries, it is typical that  the maximum charging rate is less than the maximum discharging rate \cite{website:SONY,website:LG}.}.
Let $P(t)$ denote the transmit power drawn from the battery at time slot $t$ for data transmission, which is determined at each time slot $t$ and remains unchanged during the time slot. It is bounded by
\begin{align}\label{p}
0\le P(t) \le P_{\max}, \ \forall t.
\end{align}
In each time slot $t$, energy is harvested into the battery and power is drawn from the battery for transmission. The dynamics of SOB $E_b(t)$ over time slots is given by
\begin{align}\label{sob}
E_b(t+1)=E_b(t)-\Delta tP(t)+E _s(t)
\end{align}
where by constraint \eqref{batst} and dynamics of $E_b(t)$ in \eqref{sob}, $P(t)$ should satisfy
\begin{align}
\Delta tP(t) &\le E_b(t)-E_{\min}, \ \forall t. \label{pcons}
\end{align}
The harvested energy $E_s(t)$
is determined by the amount of energy arrived, available room in the battery, and the maximum charging rate as follows
\begin{align}\label{sh}
\hspace*{-.8em}E_s(t)= \min\{E_{\max}-(E_b(t)-\Delta t P(t)), E_a(t),E_{c,\max}\}.
\end{align}

\emph{Remark}: We assume perfect charging and discharging for the battery modeling. In practice, due to battery charging inefficiency,  energy loss is expected during charging and discharging. The actual stored energy is less than the charging amount and the contributed power through discharging is larger than the actual power output. Let $\rho_{c} \in (0,1]$ and $\rho_d \in [1,\infty) $ denote the charging efficiency and discharging efficiency coefficients, respectively. Considering the charging and discharging losses, the actual stored energy $E_s(t)$ is given by
 $E_s(t)= \min\{E_{\max}-(E_b(t)-\Delta t P(t)), \rho_{c} E_a(t),\rho_{c} E_{c,\max}\}$, and the actual contributed energy through discharging is $\rho_d \Delta t P(t)$. In this work, for simplicity and without loss of generality,  we assume $\rho_c=\rho_d=1$. Our developed online power control algorithm and its analysis can be straightforwardly applied to the battery model with general values of $\rho_c$ and $\rho_d$ within their respective ranges.

For the transmission over fading,  we focus on the case where both transmitter and receiver have a single antenna. In Section~\ref{subsubsec:extension}, we  extend our proposed algorithm to the case of multi-antenna transmit beamforming. We assume a slow block fading scenario, where the channel, denoted by $h(t)$,   is assumed to be constant during time slot $t$ and changes over time slots. Assuming the receiver noise is additive white Gaussian noise with zero mean and variance $\sigma_{N}^2$, we define $\gamma(t)$ as  the normalized channel gain (against receiver noise) by $\gamma(t)\triangleq{|h(t)|^2}/{\sigma^2_N}$. We assume $\gamma(t)$ is perfectly known at the transmitter at each time slot $t$. With transmit power $P(t)$, the instantaneous rate over the channel is given by $R(t)\triangleq\log\left[1+P(t)\gamma(t)\right]$.

\section{Power Control Design for Rate Maximization }\label{sec:power control}
Define the system state $\sbf(t) \triangleq [E_a(t),\gamma(t)]$. At the beginning of each time slot $t$, the transmitter observes $\sbf(t)$ and $E_b(t)$ to determine  transmit power $P(t)$ for  time slot $t$.
Our objective is to design a power control algorithm for $\{P(t)\}$ to maximize the long-term time-averaged expected  rate, while satisfying the battery operational constraints. It can be formulated as the following optimization problem
\begin{align} 
\textbf{P1}: \quad  \max_{\{P(t)\}}  &\lim_{T \to \infty} \frac{1}{T}\sum_{t=0}^{T-1} \mbbE[R(t)]\nonumber\\
\text{subject to}  & \ \eqref{p}, \eqref{sob}, \eqref{pcons} \nonumber
\end{align}
where the expectation  is taken with respect to the system state $\sbf(t)$.

Due to the randomness of energy arrival and fading,  {\bf P1} is a stochastic optimization problem that is  challenging to solve. Furthermore, constraint \eqref{pcons} depends on the SOB  $E_b(t)$, which has time-coupling dynamics over time as shown in \eqref{sob}. This results in  power control decisions $\{P(t)\}$ being correlated over time.  If  random processes $\{\gamma(t) \}$ and $\{E_a(t)\}$ are Markovian and their statistics are all known,  it is possible to solve {\bf P1} through Dynamic Programming \cite{Bertsekas7}. However, this approach typically faces the curse of dimensionality in computational complexity to provide a practical solution. Furthermore, in practice, the statistical information of $\{\gamma(t) \}$ and $\{E_a(t)\}$, especially the energy arrival process $\{E_a(t)\}$, is difficult to obtain ahead of time, making such an assumption less realistic.

In this work, we aim to develop an online power control algorithm without relying on the statistical knowledge of $\{\gamma(t)\}$  and $\{E_a(t)$\}. In particular, we apply  Lyapunov optimization framework \cite{Neely10} to design an online  (sub-optimal) power control solution to {\bf P1}.
Under Lyapunov optimization, certain time-averaged constraints can be transformed into queue stability constraints and further be utilized to provide an online optimization solution. However, the transmit power constraint \eqref{pcons} on $P(t)$ is per time slot, resulting in time-coupled decision.  Thus, to employ Lyapunov optimization, we  first relax the per time slot constraint to a long-term time-averaged relation between $E_b(t)$, $E_s(t)$ and $P(t)$.
\subsection{Problem Relaxation}
Define the following long-term time-averaged quantities: $\bar{E}_s \triangleq \lim_{T \to \infty} \frac{1}{T}\sum_{t=0}^{T-1}\mathbb{E}[E_s(t)]$ and $\bar{P}\triangleq \lim_{T \to \infty} \frac{1}{T}\sum_{t=0}^{T-1} \mathbb{E}[P(t)]$.
We have the following long-term time-averaged  relation
\begin{align}\label{prelax}
\bar{E}_s - \Delta t\bar{P}=0.
\end{align}
To see this, note that from \eqref{sob}, the battery energy level over time $T$ has the following relation
\begin{align}\label{pr}
\mathbb{E}[E_b(T)]-\mathbb{E}[E_b(0)]=\sum_{t=0}^{T-1}\mathbb{E}[E_s(t)-\Delta tP(t)].
\end{align}
By constraint \eqref{batst}, the left hand side (LHS) is bounded. Dividing both sides of \eqref{pr} by $T$ and taking the limit $T\rightarrow \infty $, we have \eqref{prelax}. The relation in \eqref{prelax} is intuitive. It indicates that over the long run, the average energy harvested should be equal to the average energy used from the battery for transmission.

Now, replacing per-slot constraint \eqref{sob} by the long-term time-averaged constraint \eqref{prelax}, and removing battery capacity constraint \eqref{batst}, we relax the optimization problem \textbf{P1} to the following problem
\begin{align} 
\textbf{P2}:\quad\max_{\{P(t)\}} \ & \lim_{T \to \infty} \frac{1}{T}\sum_{t=0}^{T-1} \mathbb{E}[R(t)] \nonumber \\
\text{subject to} \quad
 &\eqref{p},\eqref{prelax} \nn
\end{align}
where the dependency of power control decision $P(t)$ on $E_b(t)$ in constraint \eqref{pcons} is removed.
It can be easily verified that any feasible solution to {\bf P1} is also feasible to {\bf P2}, but not vise versa. Thus, {\bf P2} is indeed a relaxed problem of {\bf P1}.

With the knowledge of only current system state $\sbf(t)$, {\bf P2} is still challenging to solve. However, the relaxation enables us to employ  Lyapunov optimization framework to develop an online power control algorithm to solve \textbf{P2}. In the following, we develop our online algorithm. Furthermore, we will show that  by our design, our proposed solution is feasible to the original problem {\bf P1}.

\subsection{Online Power Control  via Lyapunov Optimization}\label{subsec:control_policy}

We now develop an online power control algorithm to solve {\bf P2.} Based on Lyapunov optimization \cite{Neely10}, we introduce a virtual queue $X(t)$ for the SOB $E_b(t)$ as
 \begin{align}\label{vx}
X(t)= E_b(t)-A
\end{align}
where $A$ is a time-independent constant. It can be shown \cite{Neely10} that keeping the stability of the queue $X(t)$ is equivalent to satisfying constraint \eqref{prelax}.
 We will later determine the value of $A$ to ensure the proposed solution is feasible to {\bf P1}.

Since $X(t)$\ is a shifted version of $E_b(t)$, by \eqref{sob}, the queuing dynamics of $X(t)$ is given by
\begin{align} \label{dx}
X(t+1)=X(t)-\Delta tP(t)+E_s(t).
\end{align}
Note that, although $E_b(t)\ge 0$, the value of $X(t)$ can be negative.
%

Define the quadratic Lyapunov function  as $L(X(t))\triangleq X^2(t)/2.$ Define the per-slot Lyapunov drift, conditioned on $X(t)$ at time slot $t$  by
\begin{align}\label{lypdrift}
\Delta (X(t))&\triangleq \mathbb{E}\left[L(X(t+1))-L(X(t))|X(t)\right]
\end{align}
where the expectation is taken with respect to the random system state $\sbf(t)$, given queue length $X(t)$. By Lyapunov optimization framework, instead of directly using the objective in {\bf P2},  we consider the minimization of a \emph{drift-plus-cost} metric, a technique  to stabilize a queue while  optimizing the time-averaged  objective  function. The drift-plus-cost metric is defined by
\[
\Delta (X(t))+V\mathbb{E}[-R(t)|X(t)]
\]
which is a weighted sum of the per-slot Lyapunov drift
$\Delta (X(t))$ and the cost function (\ie negative of the rate) conditioned on $X(t)$ with $V>0$ being the weight.

 We first  provide an upper bound on the drift-plus-cost metric in the following lemma.
\begin{lemma}\label{lemma1}
Under any control algorithm and for any values of $X(t)$ and $V\ge0$, the drift-plus-cost expression has the following upper bound
\begin{align}
&\Delta (X(t))-V\mbbE[R(t)|X(t)] \leq\nonumber\\
& B+X(t)\mbbE[E_s(t)-\Delta tP(t)|X(t)]-V\mbbE[R(t)|X(t)] \label{ly5f}
\end{align}
where $B\triangleq \max\{E_{c,\max},\ \Delta tP_{\max}\}^2/2$.
\end{lemma}
\IEEEproof
See Appendix~\ref{app:lemma1}.

Due to the dynamics involved in $\Delta X(t)$, minimizing the drift-plus-cost metric directly is still difficult. Instead, we consider minimizing its upper bound in \eqref{ly5f}.
Specifically, we develop an online algorithm to determine $P(t)$, by minimizing the  upper bound of the drift-plus-penalty in \eqref{ly5f} in a \emph{per-slot} fashion.   That is, given $E_a(t),\gamma(t)$ and $X(t)$, taking the per-slot version of the upper bound in \eqref{ly5f} by removing the expectation $\mathbb{E}[\cdot]$ and removing the constant $B$, we have the following equivalent per-slot optimization problem
\begin{align}
\textbf{P3}:\min_{P(t)} &\quad X(t)[E_s(t)-\Delta tP(t)]-V\log\left(1+P(t)\gamma(t)\right) \nonumber\\
\text{subject to} &\quad  \eqref{p}.  \nonumber
\end{align}
Since the objective in {\bf P3} is convex and the constraint is linear in $P(t)$, {\bf P3} is a convex optimization problem and can be solved analytically. We obtain the optimal power $P^*(t)$ in closed-form as follows.
\begin{proposition}\label{prop0}
The optimal transmit power $P^*(t)$ for {\bf P3} is given by
\begin{align}\label{p**}
&\hspace*{-.5em}P^*(t) = \nn\\
&\hspace*{-.5em} \begin{cases}
P_{\max} & \textrm{for} \ X(t)>\frac{-V}{\Delta t(P_{\max}+\frac{1}{\gamma(t)}) }\\
 \frac{-V}{\Delta t X(t)}-\frac{1}{\gamma(t)} & \textrm{for} \ \frac{-V \gamma(t)}{\Delta t}\leq X(t)\leq \frac{-V}{\Delta t(P_{\max}+\frac{1}{\gamma(t)})}\\
0 & \textrm{for} \ X(t)<\frac{-V\gamma(t)}{\Delta t }.
\end{cases}
\end{align}
\end{proposition}
\IEEEproof See Appendix~\ref{app:prop0}.

Thus, at each time slot $t$, the transmitter observes the system state $\sbf(t)$ and determines transmit power $P^*(t)$ using \eqref{p**}. It then updates  $X(t)$ according to \eqref{dx}. Note that determining $P^*(t)$ does not require any statistical information of the energy arrival $E_a(t)$ or channel gain $\gamma(t)$.

As mentioned earlier, since {\bf P2} is the relaxed problem, its solution may not be feasible to {\bf P1}.   To ensure the solution $P^*(t)$ of {\bf P3} is feasible to {\bf P1}, we need to guarantee  SOB $E_b(t)$ satisfies the battery capacity constraint \eqref{batst}. Recall that two parameters $A$ and $V$ are introduced in developing the online power solution $P^*(t)$ for {\bf P3}. We will design the values of $A$ and $V$ to ensure the feasibility. However, the challenge to do so is that the normalized channel gain $\gamma(t)$  is unbounded in general for a fading channel.
This prevents us to properly design $A$ and $V$.
To provide our online algorithm feasible to {\bf P1}, in the following, we first consider the case where the fading channel gain  is upper-bounded and  derive our feasible solution. Then, we extend the solution to the case where the fading channel gain distribution has unbounded support.

\subsection{Algorithm for Fading with Bounded Channel Gain} \label{subsubsec:gmax}

We first assume the channel gain $|h(t)|^2$ is upper-bounded. Consequently, the normalized channel gain $\gamma(t)$ is upper-bounded as $\gamma(t)\le \gamma_{\max}$, where $\gamma_{\max}$ denotes the maximum gain.

As mentioned at the beginning of Section~\ref{subsec:control_policy}, maintaining the stability of $X(t)$ is equivalent to satisfying constraint \eqref{prelax}. The following lemma provides the upper and lower bounds of the virtual queue $X(t)$.
Define $\zeta_{\max}\triangleq {\gamma_{\max}}/{\Delta  t}$.
\begin{lemma}\label{lemma2}
With the proposed power control solution $P^*(t)$ in \eqref{p**}, the virtual queue $X(t)$ is bounded  for all $t$ as follows
\begin{align}\label{xbound}
X_{\text{low}}\leq X(t)\leq X_{\text{up}}
\end{align}
where $ X_{\text{low}}=-V\zeta_{\max}-\Delta t P_{\max}$ and $X_{\text{up}}= E_{c,\max}$.
\end{lemma}
\IEEEproof See Appendix \ref{app:lemma2}.

With Lemma~\ref{lemma2}, the following proposition provides the conditions of the shift constant $A$ in \eqref{vx} and the weight $V$ for which the solution $P^*(t)$ is feasible to {\bf P1}.
\begin{proposition}\label{prop1}
Assume $\gamma(t) \le \gamma_{\max}$, $\forall t$. With the proposed online power control solution $P^*(t)$ in \eqref{p**}, if $A$ in \eqref{vx} is set as
\begin{align}\label{A}
A = \Delta t P_{\max}+E_{\min}+V\zeta_{\max} \end{align}
and $V\in (0, V_{\max}] $ with
\begin{align}\label{vmax}
 V_{\max}= \frac{E_{\max}-E_{\min}-E_{c,\max}-\Delta t P_{\max}}{\zeta_{\max}},
\end{align}
then $E_b(t)$ satisfies battery capacity constraint \eqref{batst}, and the power solutions $\{P^*(t)\}$ provided by \eqref{p**} are feasible to {\bf P1}.
\end{proposition}
\IEEEproof
See Appendix~\ref{app:prop1}.

From Proposition~\ref{prop1}, substituting the expression of $A$ in \eqref{A} into \eqref{vx}, we obtain the power solution $P^*(t)$ as a function of the SOB $E_b(t) $ shown in \eqref{p**_Eb} at the top of next page,
\begin{figure*}[!t]
\begin{align}\label{p**_Eb}
P^*(t) = & \begin{cases}
0 & \textrm{for} \ E_b(t)< E_{b,\text{th1}}(t). \\
 \frac{V}{\Delta t(V\zeta_{\max}+\Delta tP_{\max}+E_{\min}-E_b(t))} -\frac{1}{\gamma(t)}
&\textrm{for}\ E_{b,\text{th1}}(t) \le E_b(t) \le E_{b,\text{th2}}(t) \\
P_{\max} & \textrm{for} \ E_b(t)>E_{b,\text{th2}}(t)
\end{cases}
\end{align}
\hrulefill
\end{figure*}
where $E_{b,\text{th1}}(t)$ and $E_{b,\text{th2}}(t)$
are two time-dependent thresholds on the battery energy level, defined by
\begin{align}
E_{b,\text{th1}}(t)&\triangleq \Delta t P_{\max}+E_{\min}+ V\left(\zeta_{\max}-\frac{\gamma(t)}{\Delta t}\right)\label{Ebth1} \\
E_{b,\text{th2}}(t)&\triangleq  \Delta t P_{\max}+ E_{\min} \nn \\
& \quad +V\left(\zeta_{\max}-\frac{\gamma(t)}{\Delta t(P_{\max}\gamma(t)+1)}\right).\label{Ebth2}
\end{align}

We summarize our proposed online power control algorithm in Algorithm~\ref{algm1}. In addition, we provide the following remarks.

\emph{Remark 1}: We see from \eqref{p**_Eb} that the solution $P^*(t)$ in \eqref{p**_Eb} is a three-stage solution depending on $E_b(t)$ of the battery: 1) When $E_b(t)$ is lower than a certain level, the transmitter stops transmission to conserve energy for future transmission; 2) When $E_b(t)$ is sufficiently high, the maximum transmit power is used for transmission; 3) When $E_b(t)$ is between the above two energy levels, \ie the battery energy level is moderate, the transmit power is set between $0$ and $P_{\max}$, depending on the current $E_b(t)$ and fading condition $\gamma(t)$.

\emph{Remark 2}: In determining $P^*(t)$, the two thresholds for $E_b(t)$  depend on the normalized channel gain $\gamma(t)$ at the current time slot $t$. In particular, a higher value of $\gamma(t)$ (\ie good channel condition) results in lower threshold values $E_{b,\text{th1}}(t)$ and $E_{b,\text{th2}}(t)$ on the energy level and higher $P^*(t)$ for data transmission. On the other hand, when the channel condition is bad, the transmitter tends to conserve energy and use less power for transmission. Thus, we see that under the proposed power control algorithm, the transmission  is carried out in an \emph{opportunistic} fashion based on the channel condition.  In particular, for a given $E_b(t)$ that is between the two thresholds,  $P^*(t)$ in \eqref{p**_Eb} resembles the water-filling power control strategy, where more power is allocated  for a better channel condition.

To clearly demonstrate the above, consider the case when $V=V_{\max}$. The two thresholds $E_{b,\text{th1}(t)}$ and $E_{b,\text{th2}}(t)$ in \eqref{Ebth1} and \eqref{Ebth2}  are respectively given by
\begin{align}
E_{b,\text{th1}}(t)&=E_{\max}-E_{c,\max} \nn\\
&-\frac{\gamma(t)}{\gamma_{\max}}(E_{\max}-E_{\min}-E_{c,\max}-\Delta t P_{\max})\label{Ebth2_Vmax} \\
E_{b,\text{th2}}(t)&=  E_{\max}-E_{c,\max}-\frac{\gamma(t)}{\Delta t(P_{\max}\gamma(t)+1)} \nn\\& \quad\cdot\left(E_{\max}-E_{\min}-E_{c,\max}-\Delta t P_{\max}\right). \label{Ebth1_Vmax}
\end{align}
We see that $E_{{b,\text{th1}}}(t)$ is a decreasing function of\  $\gamma(t)$.  For $E_{b,\text{th2}}(t)$, if $\gamma(t) \gg 1/P_{\max}$, then  $E_{b,\text{th2}}(t)$ is roughly constant with respect to $\gamma(t)$.  The power allocation $P^*(t)$ for $E_{b,\text{th1}}(t) \le E_b(t) \le E_{b,\text{th2}}(t)$ is given by
\begin{align}\label{P*vmax}
P^*(t)= \frac{V_{\max}}{\Delta t(E_{\max}-E_{c,\max}-E_b(t))} -\frac{1}{\gamma(t)}.
\end{align}
It is clear that $P^*(t)$ depends on the channel condition $\gamma(t),$\ and the ``water line" depends on the current battery energy level $E_b(t)$.
Note that the consideration of $V=V_{\max}$ is not a random choice. In Section~\ref{sec:performance analysis}, we will show that for the best performance, we should set $V=V_{\max}$.

\emph{Remark 3}: Since $V>0$,  $V_{\max}$ in \eqref{vmax} should be positive. This means the battery energy storage capacity $E_{\max}-E_{\min}$ should be larger than the sum of maximum charging and discharging amount per slot $E_{c,\max}+\Delta tP_{\max}$. This assumption generally holds for the typical battery size and usage.

\begin{algorithm}[t]
\caption{Online Transmit Power Control Algorithm under Energy Harvesting ($\gamma(t)\le \gamma_{\max}$)}\label{algm1}
Set $V\in (0,V_{\max}]$ with $V_{\max}$ given in \eqref{vmax}.\\
At time slot $t$:
\begin{algorithmic}[1]
\State Observe the system state $\sbf(t)$.
\State Solve {\bf P3} to obtain $P^*(t)$ as in \eqref{p**_Eb}.
\State Output transmit power solution: $P^*(t)$.
\end{algorithmic}
\end{algorithm}

\subsection{Algorithm for Fading with Unbounded Channel Gain}\label{subsubsec:realch}
Now, we consider a more general fading scenario where the channel gain distribution has  unbounded support (\eg  Rayleigh fading). The normalized channel gain is unbounded, \ie $\gamma(t)<\infty$. To deal with this case, we now develop a modified algorithm from Algorithm~\ref{algm1} to provide a feasible power solution for the case when $\gamma(t)>\gamma_{\max}$.

Define $\Ac \triangleq[0, \gamma_{\max} ]$ and $\Ac^c=(\gamma_{\max},\infty)$. We define the case $\gamma(t)\in \Ac^c$ as an outage event. Let $\eta$ denote the outage probability, \ie $\textrm{Prob}(\gamma(t)\in \Ac^c)=\eta$. When $\gamma(t)\in \Ac$, Algorithm~\ref{algm1} still provides the feasible solution $P(t)$ to {\bf P1}. When $\gamma(t)\in \Ac^c$, however,  constraint \eqref{pcons} may be violated, and $P(t)$ in \eqref{p**_Eb} may not be feasible. In this case, we propose the following scheme to determine $P(t)$.

Define  $E_b^e(t)=E_b(t)-\Delta t P(t)$ as the SOB at the end of time  slot $t$. Define $\bar E_b^e(t)\triangleq\frac{1}{t}\sum_{\tau=1}^{t}E_b^e(\tau)$ as the time-averaged $E_b^{e}(t)$ up to time slot $t$. For $\gamma(t)\in \Ac^c$ and $P^*(t)$ in \eqref{p**_Eb} not satisfying constraint \eqref{pcons}, we set the  transmit power as
\begin{align} \label{p_outage}
P^s(t)=\left[\frac{E_b(t)-\bar E_b^e(t-1)}{\Delta t}\right]^+
\end{align}
where $[x]^+ \triangleq \max(x,0)$.

\emph{Remark:} The main idea of our scheme is that  we use the time-averaged  battery energy level $E_b(t)$ from the past to determine $P(t)$, so that at the end of the time slot, the battery energy level remains at its historical  time-averaged level.
This idea comes from the observation that in the case of $\gamma(t)\in\Ac$, our proposed algorithm under Lyapunov optimization tries to maintain the SOB $E_b(t)$ at a certain level. Thus, when the outage event occurs temporarily, we control the transmission power such that $E_b(t)$ is still roughly being maintained at its historical level as in the non-outage case. As a result, the battery energy dynamics over time will not be disturbed due to the outage event.

We summarize our online transmit power control algorithm for the general unbounded fading  case in Algorithm~\ref{algm2}. As discussed earlier, there are two main benefits provided by our  proposed algorithm to improve the long-term time-averaged rate: 1) Strategic energy conservation through energy control in the battery; 2) Opportunistic transmission through power control over fading. As we will see in simulation results in Section~\ref{sec:simulation results}, these benefits are evident in improving the transmission data rate.
\begin{algorithm}[t]
\caption{Online Transmit Power Control Algorithm under Energy Harvesting ($\gamma(t) <\infty$)}\label{algm2}
Choose $\eta$. Determine $\gamma_{\max}$ from $\eta$.

 At time slot $t$:
\begin{algorithmic}[1]
\State Observe the system state $\sbf(t)$.
\State Apply Algorithm 1 to produce $P^*(t)$. Set $P^s(t)=P^*(t)$.
\If {$\gamma(t)\in \Ac^c$ and $P^s(t)>(E_b(t)-E_{\min})/{\Delta t}$\ }
 obtain $P^s(t)$ as in \eqref{p_outage}.
\EndIf
\State Update $E^e_b(t) = E_b(t)-\Delta tP^s(t)$.
\State Update $\bar{E}^e_b(t) = \frac{1}{t} \left[(t-1) \bar{E}^e_b(t-1)+E^e_b(t)\right]$.
\State Output the transmit power solution $P^s(t)$.
\end{algorithmic}
\end{algorithm}
\subsection{Extension to Multi-antenna Beamforming Scenarios}\label{subsubsec:extension}
In the above, we have focused on the single-antenna case. Our proposed algorithm can be easily extended to the scenarios of multi-antenna beamforming.

For example, consider a MISO system with $N$ transmit antennas and a single receive antenna. Under the block fading model, the channel  vector between the transmitter and the receiver at time slot $t$ is denoted by $\hbf(t)=[h_1(t),\ldots,h_N(t)]^T$. With perfect knowledge of $\hbf(t)$ at the transmitter and the optimal transmit beamforming, the normalized channel gain at time slot $t$ is  given by $\gamma(t)\triangleq{||\hbf(t)||^2}/{\sigma_N^2}$.
The instantaneous rate over the channel during time slot $t$ has the same expression as we consider before:
$ R(t)=\log\left(1+P(t)\gamma(t)\right)$. Thus, the only difference is about channel gain $\gamma(t)$ and its distribution. Our proposed online algorithm (Algorithm \ref{algm2}) can be directly applied for this transmit beamforming scenario.

Similarly, the algorithm can be applied for the  single-input multi-output (SIMO) case with receive beamforming, 、or MIMO beamforming. For the latter, transmit and receive beam vectors are selected as the principle right and left singular vector of the MIMO channel, denoted by $\Hbf(t)$. In this case, the effective normalized channel gain is $\gamma(t) = \sigma_1^2(t)/\sigma_N^2$, where $\sigma_1^2(t)$ is the largest singular value of $\Hbf(t)$.  The expression of instantaneous rate $R(t)$ is still the same as before.

\section{Performance Analysis }\label{sec:performance analysis}
In this section, we analyze the performance of our proposed online power control algorithms.
\subsection{Bounded Fading Scenario}
We first consider the case where $\gamma(t)\in\Ac, \forall t$, and analyze the performance of Algorithm~\ref{algm1}. Let $\bar{R}^s(V,\Ac)$ denote the achieved objective value of {\bf P1} under Algorithm~\ref{algm1}. Let  $\bar{R}^{\textrm{opt}}(\Ac)$  denote the maximum objective value of {\bf P1} under the optimal solution.
The following theorem provides a bound of the performance of Algorithm~\ref{algm1} to  $\bar{R}^{\textrm{opt}}(\Ac)$.

\begin{theorem}
 \label{thrm1}
 Assume $\gamma(t)\in \Ac$, $\forall t$. Assume the system state $\sbf(t)$ are i.i.d over time. Under Algorithm \ref{algm1}, the performance is bounded from the maximum value $\bar{R}^{\textrm{opt}}(\Ac)$ of \textbf{P1} by
\begin{align}\label{alg_ub1}
\bar{R}^{\textrm{opt}}(\Ac)-\bar{R}^s(V,\Ac)&\leq \frac{B}{V}
\end{align}
where $B$ is defined below \eqref{ly5f}.
\end{theorem}
\IEEEproof
See Appendix \ref{app:thrm1}~.

We have the following remarks on Theorem~\ref{thrm1}.

\emph{Remark 1}: Theorem~\ref{thrm1} provides an  upper bound on the gap of the long-term time-averaged rate  of our proposed algorithm  away from  $\bar{R}^\textrm{opt}(\Ac)$ by the optimal solution. It is in the order of $\Oc(1/V)$. Thus, larger $V$ is desirable. However, due to battery capacity constraint, by Proposition~\ref{prop1}, $V$ has to be chosen within $(0,V_{\max}]$. Thus, to minimize the performance gap, we should always chose $V=V_{\max}$.

\emph{Remark 2}: For the upper bound in \eqref{alg_ub1}, note that  $B$ is  only related to the  battery maximum charging and discharging rates, not the battery capacity, while $V_{\max}$ in \eqref{vmax} increases with battery capacity.  Thus, Algorithm~\ref{algm1} provides an asymptotically optimal solution for {\bf P1}, as the battery storage capacity $(E_{\max}-E_{\min})$ increases .

\emph{Remark 3}: Although the upper bound in \eqref{alg_ub1} is provided under the i.i.d. assumption, the system state $\sbf(t)$ can be relaxed to accommodate the case where $\sbf(t)$ evolving in ergodic non-i.i.d. fashion.  Specifically, if both normalized channel gain $\{\gamma(t)\}$ and energy arrival $\{E_a(t)\}$ processes are modeled as the finite state Markov chains, we can show a similar bound (\ie $\Oc(1/V)$) under Algorithm~\ref{algm1},  by applying a multi-slot Lyapunov drift technique \cite{Neely10}. We omit details for brevity.

\subsection{Unbounded Fading Scenario}
With probability $\eta$,  $\gamma(t)\in \Ac^c$. In this case, the outage event occurs, and power solution is determined differently.  Let $\bar{R}^\textrm{opt}(\Ac^c)$ denote the maximum objective value of  {\bf P1} under the optimal solution and $\bar{R}^s(\Ac^c)$ denote the achieved objective using $P(t)$ in \eqref{p_outage}, both in the presence of  the outage. The following lemma provides an upper bound on the performance when the outage occurs.

\begin{lemma}\label{lemma3}
Assume that system state $\sbf(t)$ is i.i.d over time, and the channel has a normalized channel gain distribution $f(\gamma)$. For  $\gamma(t)\in{\cal A}^c$, under Algorithm~\ref{algm2},  the performance is bounded by
\begin{align}\label{alg_ub2}
\bar R^\textrm{opt}(\Ac^c)-\bar{R}^s(\Ac^c)\leq G
\end{align}
where constant $0<G<\infty$ is a function of $f(\gamma)$ and $\gamma_{\max}$.
\end{lemma}
\IEEEproof
See Appendix~\ref{app:lemma3}~.

As indicated in Lemma~\ref{lemma3}, the upper bound $G$ can be obtained for any specific channel distribution. In particular, for SIMO or MISO beamforming with channel vector $\hbf(t)$, assume Rayleigh fading, \ie element $h_n(t)$ in   $\hbf(t)$ is complex Gaussian with zero mean and variance $\sigma_h^2$, for $n=1,\ldots,N$. By Lemma~\ref{lemma3}, we obtain the expression of $G$ in the following corollary.
\begin{corollary}\label{cor1}
  Assume Rayleigh fading channels. Under the assumptions of Lemma~\ref{lemma3}, $G$ is given by
\begin{align}\label{G}
G&= C\int_{\frac{\gamma_{\max}}{\bar{\sigma}_h^2}}^\infty \log\left(1+\bar{\sigma}_h^2P_{\max}\gamma \right)\gamma^{N-1}e^{-\gamma}d\gamma
\end{align}
with $\bar{\sigma}_h^2\triangleq \sigma_h^2/\sigma_N^2$ and  $C\triangleq \left[\hat{\Gamma}\left(N,\frac{\gamma_{\max}}{\sigma_h^2} \right)\right]^{-1}$, where  $\hat{\Gamma}(n,y)\triangleq\int_y^{\infty} x^{n-1}e^{-x} dx$ is the upper incomplete Gamma function.
In particular, for $N=1$, we have
\begin{align}\label{G:N=1}
G=&\log\left(1+P_{\max}\gamma_{\max}\right) + e^{\gamma_o}\hat{\Gamma}(0,\gamma_o)
\end{align}
where $\gamma_o \triangleq {\gamma_{\max}}/{\bar{\sigma}_h^2}+{1}/{(\bar{\sigma}_h^2P_{\max})}$.
\end{corollary}
\IEEEproof
See Appendix~\ref{app:cor1}~.

Let $\bar{R}^\textrm{opt}$ denote the the maximum objective value of {\bf P1}. Let $\bar{R}^s(V,\eta)$ denote the achieved objective under Algorithm~\ref{algm2}, where we emphasize the dependency  of the achieved objective value   on the control parameter $V$ and the outage probability $\eta$ used in our algorithm.    Combining the results in Theorem~\ref{thrm1} and Lemma~\ref{lemma3}, we have the following performance bound.
\begin{theorem}\label{thrm2}
Assume the system state $\sbf(t)$ is i.i.d over time.  For the fading channel with any  fading distribution, given the outage probability $\eta$, the performance under Algorithm~\ref{algm2} is bounded from $\bar{R}^{\textrm{opt}}$ by
\begin{align}\label{alg_ub}
\bar{R}^{\textrm{opt}}-\bar{R}^s(V,\eta)\leq (1-\eta)\frac{B}{V}+\eta G.
\end{align}
\end{theorem}
\IEEEproof
See Appendix~\ref{app:thrm2}.

Theorem~\ref{thrm2} provides an upper bound on the performance gap of Algorithm~\ref{algm2} to the optimal solution of {\bf P1} over a general fading scenario. The bound depends on the outage probability $\eta$ we choose. So long $\eta$ is chosen to be small, the effect due to outage on the bound will be small. As we will see in our simulation, the difference on the actual performance of our proposed algorithm under the bounded channel and unbounded channel is negligible, provided $\eta$ is small.
In Section \ref{subsec:simulation V}, we show through simulation that the performance  approaches to the optimal solution quickly as battery size increases.

Note that in the unbounded fading channel scenario, Algorithm~\ref{algm1} is used for $\gamma(t) \in \Ac$ with probability $1-\eta$. Thus, Remarks 1 and 3 after Theorem~\ref{thrm1} are also applicable to Theorem~\ref{thrm2}. However,  due to the gap $G$ in the case of outage $\gamma(t)\in \Ac^c$, as the battery capacity goes to infinity, we can only guarantee Algorithm~\ref{algm2} to asymptotically  have a bounded gap $\eta G$ in performance to the optimal solution of {\bf P1}.

\section{Simulation Results}\label{sec:simulation results}
In this section, we examine the performance of our proposed online power control algorithm. We assume that the energy arrival amount $E_a(t)$\  per slot follows a compound Poisson process with a uniform distribution. We set the default Poisson arrival rate   $\lambda=0.5 $ unit/slot.
 The
 amount of energy per unit is uniformly distributed between $[0, 2\alpha ]J$, with the default mean amount  $\alpha=0.2J$.  The battery minimum energy level is set to  $E_{\min}=0$. For the battery maximum energy level, unless specifically specified, we set the default value to $E_{\max}=50J$.  Also, the maximum charging amount per slot is  $E_{c,\max}=0.3 J$, and the maximum transmission power is $P_{\max}=0.5 W$.  We set time slot duration to be   $\Delta t=1$ sec.

By default, we consider single antenna $N=1$. We generate  channel $h(t)$ as i.i.d. complex Gaussian random variable over time $t$ with the mean normalized channel gain  $\mathbb{E}[\gamma(t)]=10 $ dB.  We set the outage probability $\eta=1\%$. This results in  $\gamma_{\max} = 16.6$ dB. For comparison purpose, we consider our proposed algorithms in two fading scenarios:
\begin{list}{\labelitemi}
{
\setlength\leftmargin{1.3em}
\setlength\labelwidth{1em}
\setlength\labelsep{.3em}
}
\item[(a)]\emph{Bounded fading $\gamma(t)\le \gamma_{\max}$}: We first generate the channel as described above. If $\gamma(t)>\gamma_{\max}$, we set $\gamma(t)=\gamma_{\max}$. We apply Algorithm~\ref{algm1} to obtain the transmit power $P(t)$.
\item[(b)] \emph{Unbounded fading}: The channel is generated as complex Gaussian as described above. We apply Algorithm~\ref{algm2} to determine the transmit power $P(t)$. For both (a) and (b), we set default $V=V_{\max}$.
\end{list}
To compare with other online power control algorithms,  note that, as discussed in Section~\ref{sec:intro}, existing online power control strategies (\cite{SinhaChaporkar:NCC12,WangTajer:JCOM12,HoZhang:TSP12,OzelYener:TSA11,Sharmaetal:TWCOM10,Jing:JWCOM09,BlascoGunduz:JWCOM13}) are either for AWGN channels only,  or based on known statistical knowledge of energy arrivals and fading channels. Also, we have a more detailed model of battery operational constraints on energy harvesting and power supply. As a result, our proposed algorithm cannot be directly compared with algorithms in \cite{SinhaChaporkar:NCC12,WangTajer:JCOM12,HoZhang:TSP12,OzelYener:TSA11,Sharmaetal:TWCOM10,Jing:JWCOM09,BlascoGunduz:JWCOM13}.   Nonetheless,  we include a heuristic online water-filling algorithm proposed in \cite{OzelYener:TSA11} for comparison, in which the fading statistics is assumed to be known to determine the transmission power\footnote{We slightly modify the solution to meet the battery operational constraints.}. Furthermore, for a fair comparison, we consider two alternative online algorithms that also only rely on the current system state without requiring its statistical information. The three algorithms are described below:
\begin{list}{\labelitemi}
{
\setlength\leftmargin{1.3em}
\setlength\labelwidth{1em}
\setlength\labelsep{.3em}
}
\item[(c)] \emph{Energy adaptive water-filling algorithm (EAWF)  \cite{OzelYener:TSA11}}: Compute a cutoff fade $\gamma_0$ at each time slot as the solution of the following equation
\begin{align}
\int_{\gamma_0}^\infty \left(\frac{1}{\gamma_0}-\frac{1}{\gamma}\right)f(\gamma)d\gamma=E_b(t).
\end{align}
Then, given $\gamma(t)$, the transmission power is determined as $P(t)=\min\left\{\left[\frac{1}{\gamma_0}-\frac{1}{\gamma(t)}\right]^+, P_{\max}, (E_b(t)-E_{\min})/\Delta t\right\}$. This algorithm exploits the channel fade and uses energy adaptive water-filling to improve the transmission rate.

\item[(d)] \emph{Greedy algorithm}: At each time slot, the transmitter uses the maximum possible power based on $E_b(t)$ to maximize the transmission rate at current time slot $t$, \ie
\begin{align*}
\max_{P(t)} R(t) \quad \text{subject to \eqref{p}, \eqref{sob}, \eqref{pcons}}
\end{align*}
which results in  $P(t)=\min \{(E_b(t)-E_{\min})/{\Delta t},P_{\max}\}$.
\item[(e)]\emph{Power halving algorithm}: At each time slot, the transmitter uses half of the maximum possible power given by the greedy algorithm in (d). Different from the greedy algorithm, this simple heuristic algorithm intends to conserve harvested energy in the battery.
\end{list}
Note that, when implementing algorithms (c)--(e), the complex Gaussian channel is used as in (b) unbounded fading case.

\subsection{Average Rate Convergence over Time}

\begin{figure}[t]
\centering
\includegraphics[scale=.5]{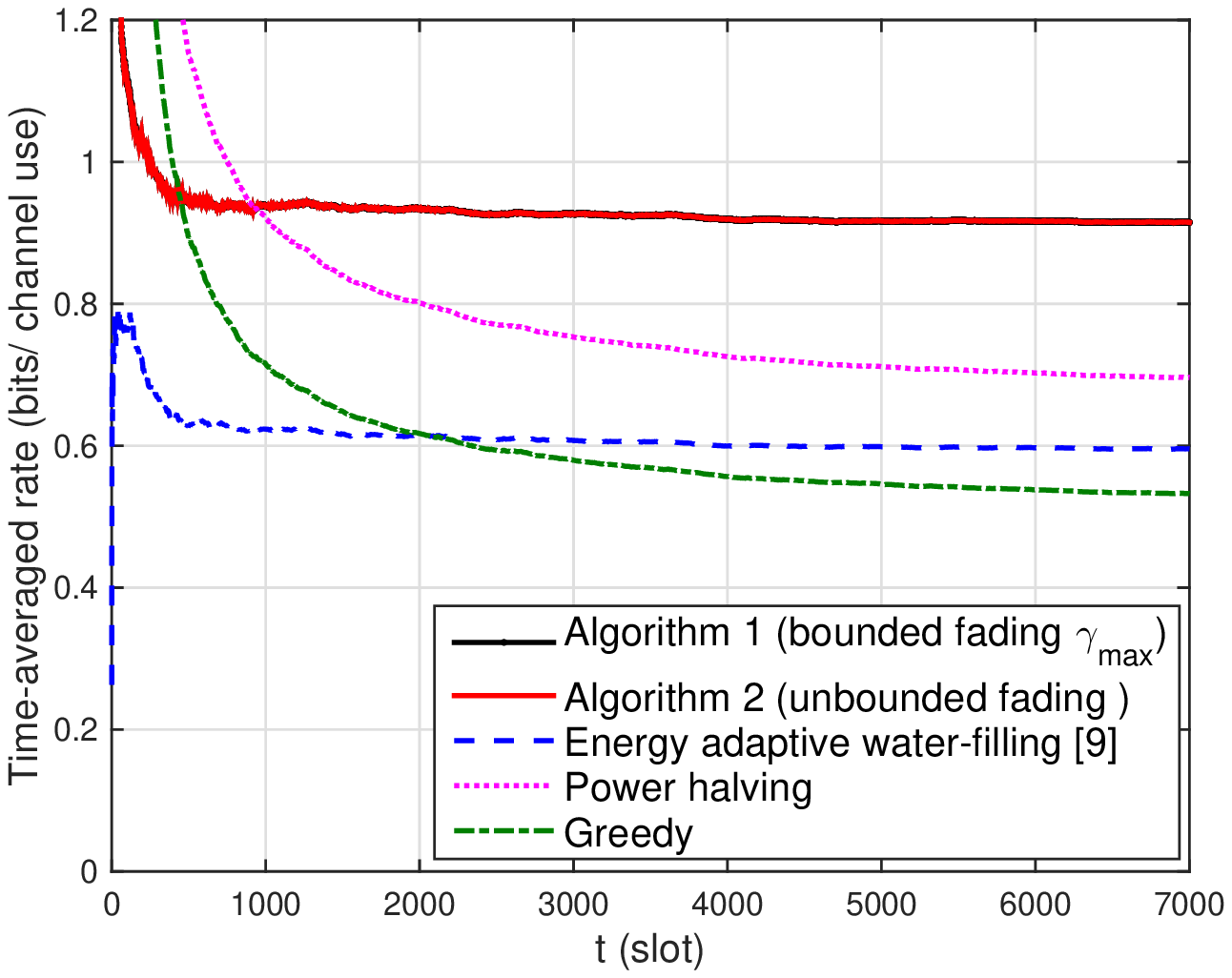}
\caption{Time-averaged rate vs. time slot $t$ ($E_{\max}=50 J, E_b(0)= E_{\max}$).
}
\label{fig.1}
\centering
\includegraphics[scale=.5]{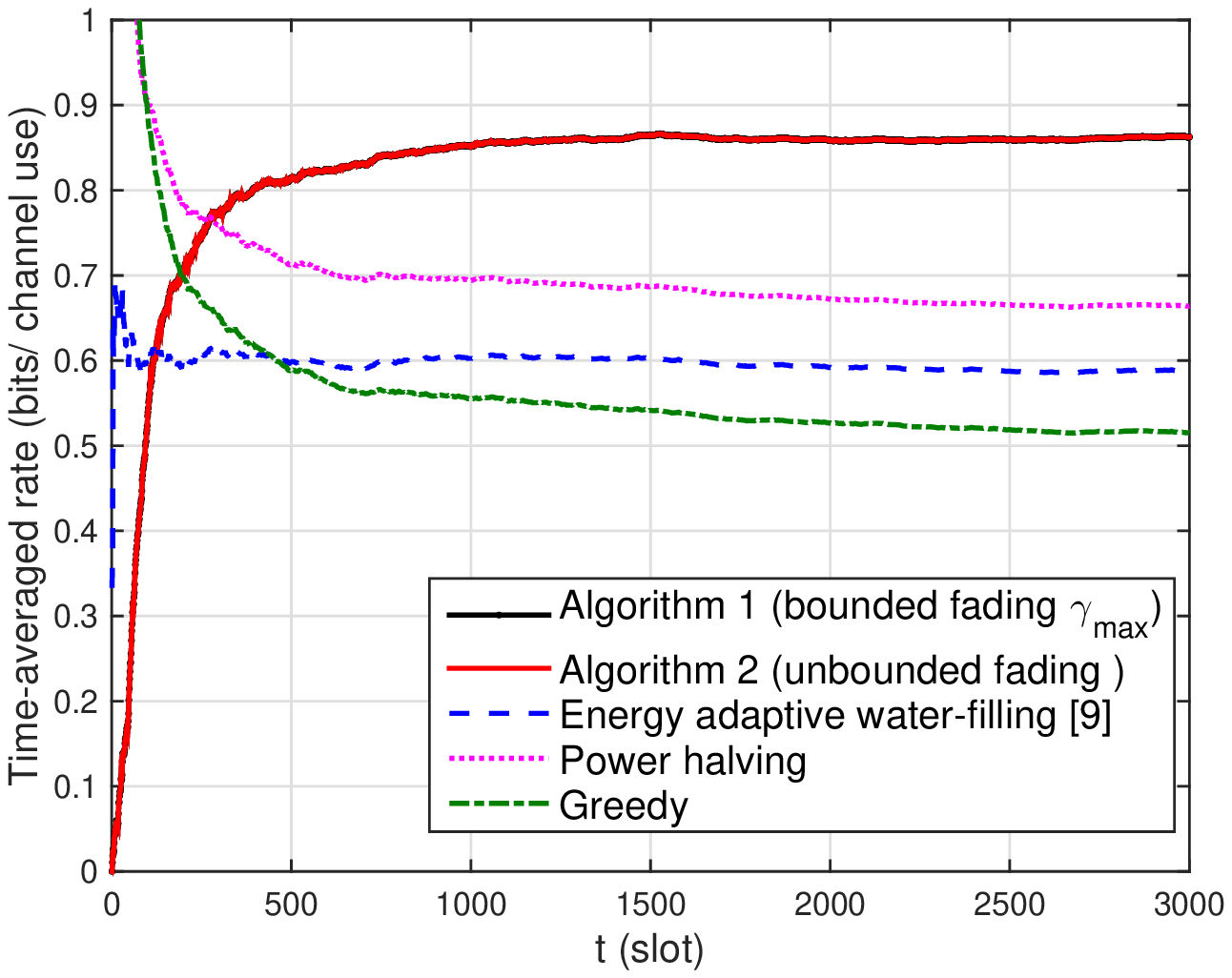}
\caption{Time-averaged rate vs. time slot $t$ ($E_{\max}= 10J, E_b(0)={ E_{\max}}/{2}$).
}
\label{fig.2}
\end{figure}

Let $R^s(t)$ denote the achieved rate at time slot $t$.
In Fig.~\ref{fig.1}, we plot the time-averaged rate $\frac{1}{T}\sum_{t=0}^{T-1} R^{s}(t)$, averaged over  Monte Carlo runs, versus time slots. We set $E_b(0)=E_{\max}=50J$. As we see, with 1\% outage probability setting, the performance of Algorithms~\ref{algm1} and \ref{algm2} under the two fading scenarios (bounded and unbounded) result in nearly identical performance. Furthermore, our proposed online power control algorithm provides significant performance improvement over all the other three algorithms (c)--(e). Specifically, the achieved average rate by Algorithm~\ref{algm2} is  about 70\% higher than that by the greedy algorithm, and about 50\% and 30\% higher than the EAWF and power halving algorithms,  respectively. As we see, even though  the EAWF algorithm in \cite{OzelYener:TSA11} assumes known fading statistics,  Algorithm~\ref{algm2} still provides more than 50\% improvement on the average rate without requiring any fading statistics. The performance gain of our proposed algorithm over these alternative algorithms comes from strategic energy conservation and opportunistic transmission.

We repeat the experiment with a smaller battery capacity with $E_{\max}=10J$.\ The initial state of battery is set to $E_b(0)=E_{\max}/2$. As shown in Figs.~\ref{fig.2}, similar performance comparisons after convergence can be observed. In addition, although not shown, we observe that for the same ratio of  initial energy level over the battery capacity (\ie $E_b(0)/(E_{\max}-E_{\min})$), the convergence time  is much shorter for a battery with smaller capacity. This behavior is intuitive since with a smaller capacity room, it takes less time slots to search for the relatively stabilized energy level for $E_b(t)$ under the same system setup.

\begin{figure}[t]
\centering
\includegraphics[scale=.5]{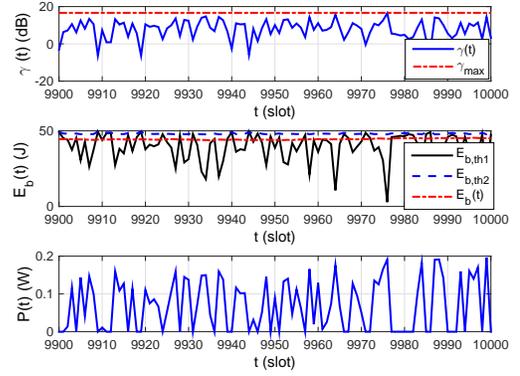}
\caption{Time trajectory of system parameters: Top:  $\gamma(t)$; Middle: $E_b(t)$, $E_{b,\text{th1}}(t)$, $E_{b,\text{th2}}(t)$; Bottom: $P(t)$. ($E_{\max}=50J$)}
\label{fig.3}
\end{figure}

\subsection{Effect of Channel Fading}\label{subsec:simulation fading}
We study the dependency of power allocation by the proposed algorithm on the fading channel. For this purpose, we consider the bounded fading scenario and Algorithm~\ref{algm1}. In Fig. \ref{fig.3}, we plot the normalized channel gain $\gamma(t)$, the SOB $E_b(t)$ and thresholds $E_{b,\text{th1}}(t)$ and $E_{b,\text{th2}}(t)$, and the allocated power $P^*(t)$ by Algorithm~1 versus time slot $t$  in the top, middle, and bottom subplots, respectively. As discussed in Remarks 3 at the end of Section \ref{subsubsec:gmax} for $V=V_{\max}$, we see that threshold  $E_{b,\text{th1}}(t)$ in \eqref{Ebth1_Vmax} changes according to  $-\gamma(t)$, while  $E_{b,\text{th2}}(t)$ is roughly the constant over time. The battery energy level $E_b(t)$ roughly maintains at a level between the two thresholds. At the bottom of Fig.~\ref{fig.3}, we see that the power $P^*(t)$ is determined approximately according to the channel condition with a higher power for a better channel gain. This demonstrates that the transmission is opportunistic based on channel quality.

\subsection{Effect of Parameter V}\label{subsec:simulation V}
We evaluate the performance of our proposed algorithms for $V \in (0,V_{\max}]$ in Fig.~\ref{fig.5}. The long-term time-averaged data rate is averaged over 100 Monte carlo runs.    We see that, under the proposed algorithms, the average data rate initially increases with $V$ sharply, and then gradually converges to a stable value. This trend is consistent with results in  Theorems~\ref{thrm1} and \ref{thrm2}, where the bounded gap to the optimal performance decreases with $V$.  Furthermore, since the averaged rate quickly converges to its stable value with a relatively small value of $V$, the value of $V_{\max}$ can be relatively small. Since $V_{\max}$ is a function of battery capacity, this indicates that a smaller battery storage capacity would be sufficient to achieve a near-optimal performance. This observation is confirmed in our next study on the battery capacity. In contrast, the other three alternative algorithms does not change with $V$, and thus the averaged rate remains flat.

\subsection{Performance vs. Battery Capacity}
\begin{figure}[t]
\centering
\includegraphics[scale=.5]{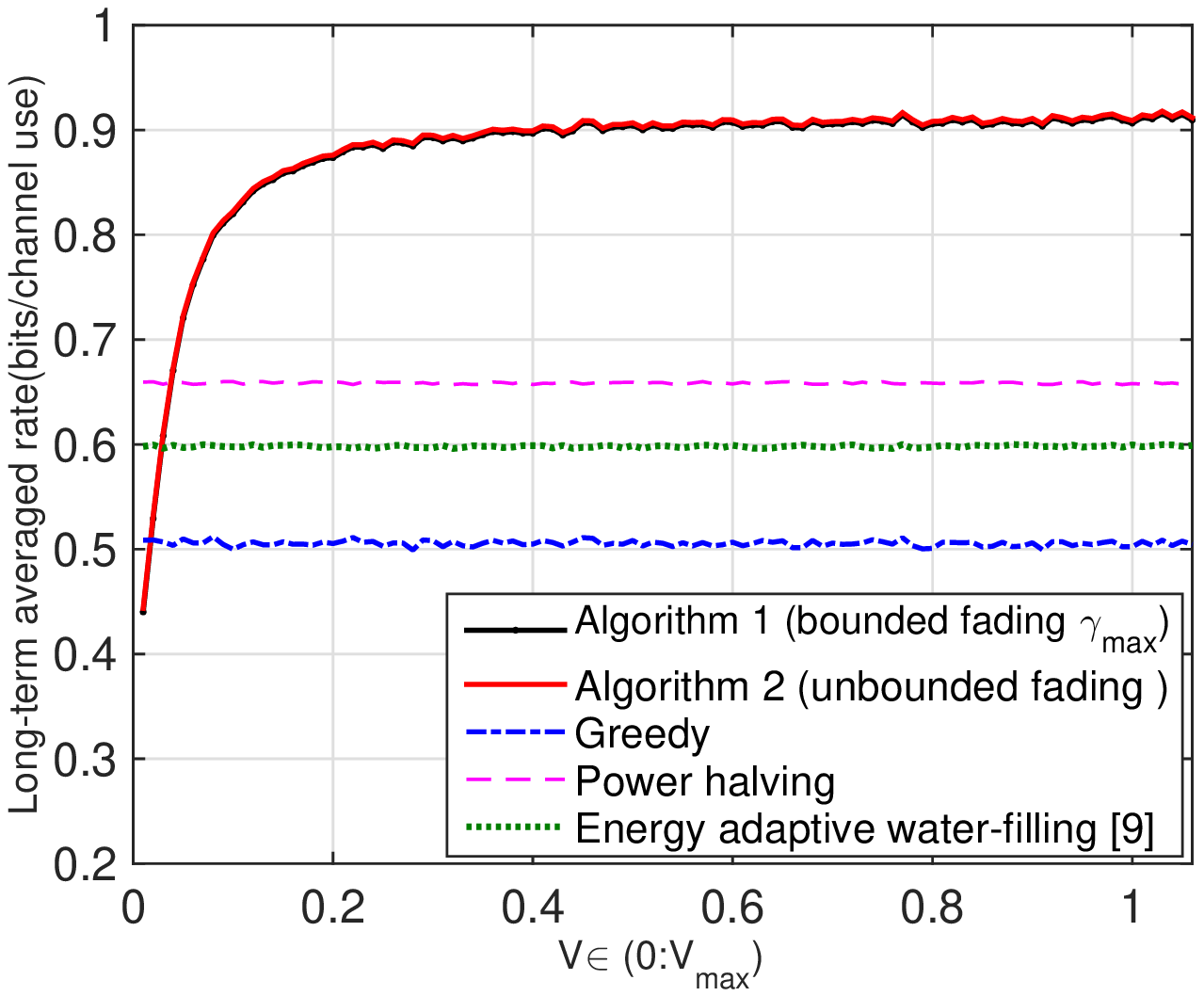}
\caption{The long-term averaged  rate vs. $V$ for $V\in (0, V_{\max}]$. }
\label{fig.5}
\includegraphics[scale=.5]{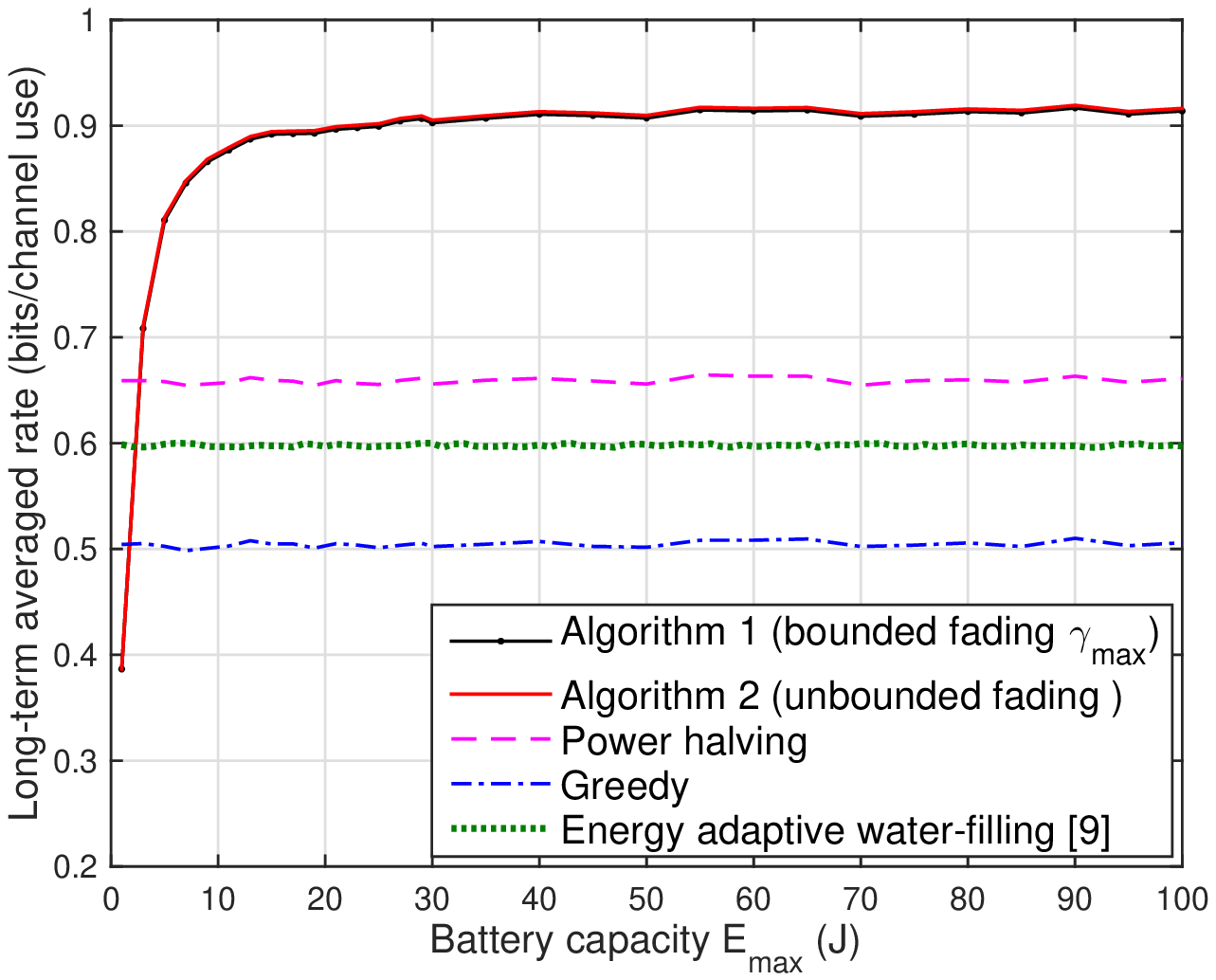}
\caption{The long-term averaged rate vs. battery size $E_{\max}$.}
\label{fig.4}
\end{figure}

In Fig.~\ref{fig.4}, we show the long-term averaged data rate under different battery capacity $E_{\max}$. We see that the performance gain over the other three alternative algorithms grows fast as the battery capacity  $E_{\max}$ increases from $1J$ to $10J$ and becomes saturated afterwards. First, this demonstrates  the effectiveness of our proposed online power control algorithm even for a small ratio of the battery capacity over the expected energy arrival rate $\alpha\lambda$. Second, we observe that under our proposed algorithms, the performance benefits significantly from a larger battery capacity, because the storage is crucial for better performance.

As the battery size continues to increase, the maximum power $P_{\max}$ and charging rate $E_{c,\max}$ become the limiting factors, and  as $V_{\max}$ increases with $E_{\max}$, the performance gradually converges to that of the optimal solution. This clearly shows that, under Algorithm~\ref{algm2}, a relatively small battery storage capacity would be sufficient for a near-optimal performance. Further increasing battery size will not be effective in improving the performance. In contrast, for the greedy algorithm, due to the greedy nature, its performance is limited by $E_{c,\max}$ and $P_{\max}$, and does not change with the battery size. The same applies to the performance of the power halving and EAWF algorithms whose performances are also unchanged with the battery size.

\subsection{Performance vs. Energy Arrival Rate $\lambda$ and $\text{SNR}$}
\begin{figure}[t]
\centering
\includegraphics[scale=.45]{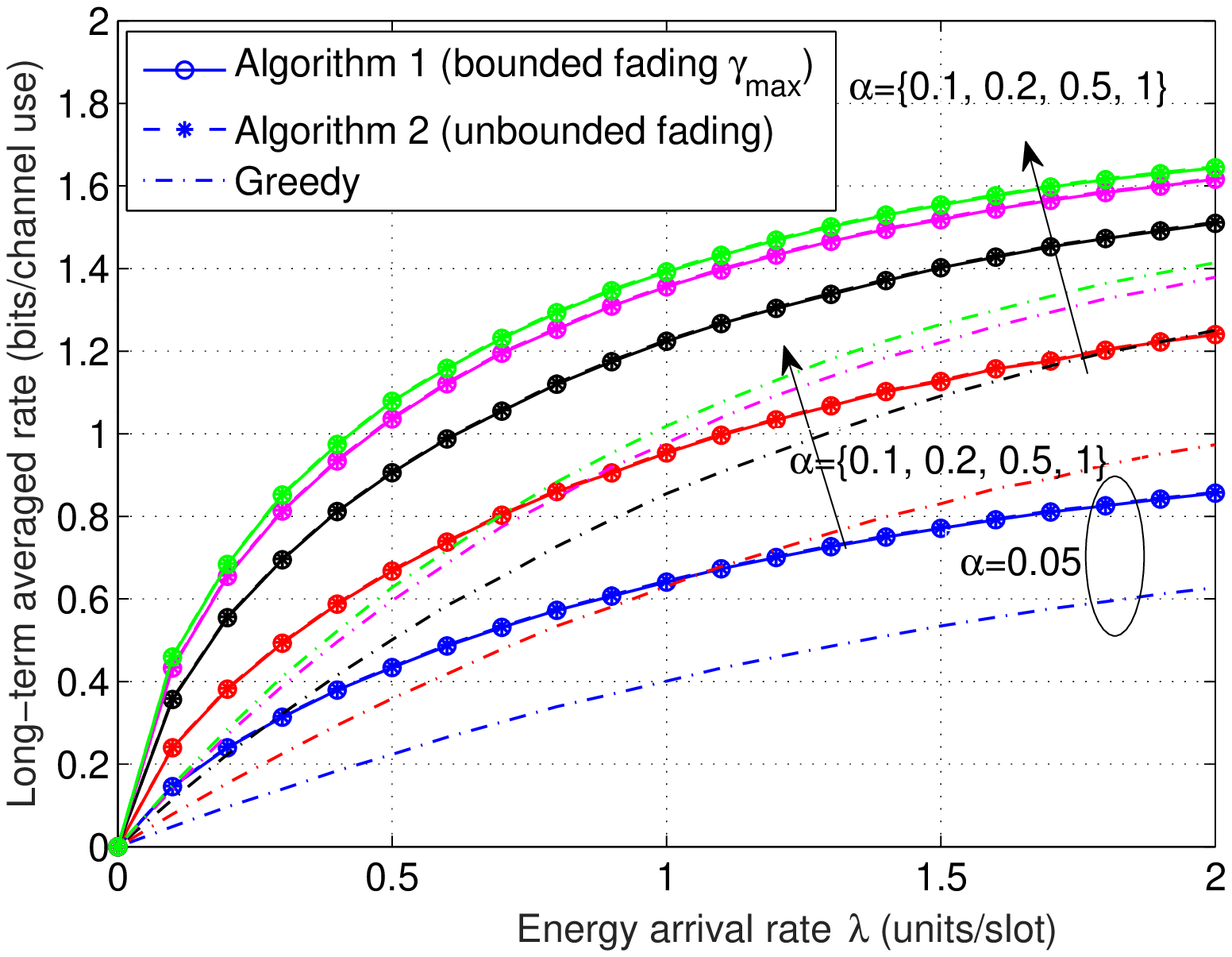}
\caption{The long-term time-averaged expected rate vs. energy arrival $\lambda$.}
\label{fig.6}
\centering
\includegraphics[scale=.55]{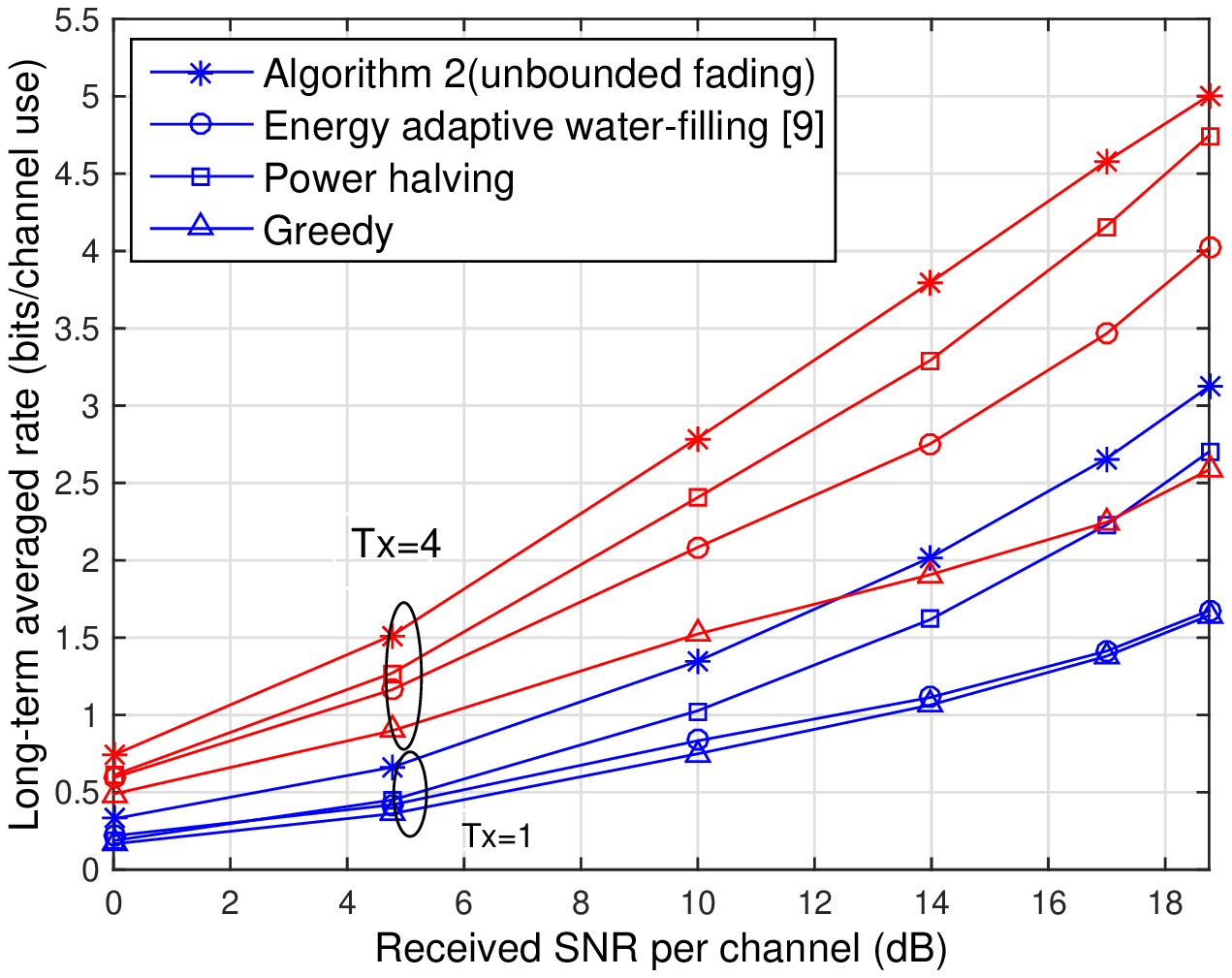}
\caption{The long-term time-averaged rate vs. received SNR per channel ($\alpha=0.1J$, $\lambda= 0.3$ unit/slot). }
\label{fig.7}
\end{figure}

In  Fig.~\ref{fig.6},  we examine the long-term average data rate under various energy arrival intensities specified by arrival rate $\lambda$ and mean arrival amount $\alpha$. The data rate monotonically increases with both $\lambda$ and $\alpha$. The rate of increment becomes smaller as $\alpha$ becomes larger. As more energy is stored in the battery, higher transmit power is used and the data rate is in the non-linear region
with respect to transmit power. Thus, less rate increment is observed.
For the comparison purpose, the performance of the greedy algorithm is also plotted. We see the gain of our proposed algorithm over the greedy algorithm is consistent over various values of $\lambda$ and $\alpha$.

Next, we evaluate the performance of Algorithm~\ref{algm2} under  MISO beamforming. Fig.~\ref{fig.7} shows the long-term time-averaged data rate versus the average received SNR per channel $\mathbb{E}[|h_n(t)|^2/\sigma_N^2]$, for the number of transmitter antennas $N=1,2,4$.  We set $\alpha = 0.1J$ and $\lambda=0.3$ unit/slot. We also include the other three algorithms (c)-(e) for comparison. As expected, the average rate increases with $N$ due to the beamforming gain, and with SNR. As we see, Algorithm ~\ref{algm2} outperforms all the other three algorithms for all values of SNR and $N$.
 In particular, the rate improvement by Algorithm~\ref{algm2} over the greedy algorithm increases significantly with both SNR and $N$. Comparing with the power halving algorithm, the rate improvement by  Algorithm~\ref{algm2} is roughly constant over SNR and $N$. This is because the power halving algorithm also attempt to conserve energy in the battery for the future use. This demonstrates the importance of controlling the stored energy in the battery for transmission over fading, especially as SNR and $N$ increases. The performance gap between  Algorithm~\ref{algm2} and the EAWF algorithm is reduced as $N$ increases. Note that as $N$ increases, the fading distribution of the effective channel  changes and shifts to higher channel gain. Since both algorithms provide water-filling like power control for opportunistic transmission,  this demonstrates the benefit of taking advantage of opportunistic transmission based on channel conditions.    


\section{Conclusion}\label{sec:conclusion}
In this paper, we have designed an online transmission power control algorithm for transmission over fading with  energy harvesting and storage devices at the transmitter for power supply. Aiming at maximizing the long-term time-averaged transmission rate under the battery operational constraints, we formulate the stochastic optimization problem for transmission power control. By developing techniques to transform the problem, we leverage Lyapunov optimization to proposed an online power control algorithm. In particular,  we develop an approach to tackle the difficulty faced in handling unbounded channel fading which otherwise cannot be dealt with directly through Lyapunov optimization. Unlike most existing online power control algorithms, our proposed algorithm only depends on the current energy arrival and channel fade condition, without requiring their statistical knowledge. In addition, our online power solution is provided in closed-form that is simple to implement. We show that our power control solution not only provides energy conservation control of the battery, but also results in an opportunistic transmission style based on fading condition, resembling a ``water-filling" like solution.
Through analysis, we show that our proposed algorithm provides a bounded performance gap  to the optimal solution for a general fading distribution.  In addition, we show that our solution applies to the general multi-antenna beamforming scenarios. Simulation studies show that our proposed online power control algorithm significantly outperforms other alternative online algorithms. \ \bibliographystyle{IEEEtran} %

\appendices
\section{Proof of Lemma~\ref{lemma1}}\label{app:lemma1}
\IEEEproof
From the dynamics of $X(t)$ in \eqref{dx}, we have
\begin{align}\label{xupsqrf}
&L(X(t+1))-L(X(t))\nn \\
&=\frac{X^2(t+1)-X^2(t)}{2} \nonumber\\
&=\;\ \frac{(E_s(t)-\Delta tP(t))^2}{2}+X(t)(E_s(t)-\Delta tP(t)).
\end{align}
From \eqref{sh}, we have $0\le E_s(t)\le E_{c,\max}$. Along with constraint \eqref{p} on $P(t)$, we have
\begin{align}\label{up}
(E_s(t)-\Delta t P(t))^2&\le \max\{E_{c,\max}, \Delta t P_{\max}\}^2.
\end{align}

Taking expectation at both sides of \eqref{xupsqrf} conditioned on $X(t)$ and considering \eqref{up}, we have the per-slot Lyapunov drift being upper bounded by
\begin{align}\label{bd}
\Delta (X(t))&=\mbbE\left[L(t+1)-L(t)|X(t)\right]\nonumber\\&\leq B+X(t)\mbbE\left[E_s(t)-\Delta tP(t)|X(t)\right]
\end{align}
where $B\triangleq \max\{E_{c,\max},\Delta tP_{\max}\}^2/2$.
Adding $-V\mbbE\left[R(t)|X(t)\right]$ to both sides of \eqref{bd}, we have the upper bound on the  drift-plus-cost metric as in \eqref{ly5f}.
\endIEEEproof

\section{Proof of Proposition~\ref{prop0}}\label{app:prop0}
\IEEEproof
Denote the objective of \textbf{P3 }as $J(P(t))$
\begin{align*}
J(P(t))\triangleq X(t)\left(E_s(t)-\Delta tP(t)\right)-V\log\left(1+\gamma(t) P(t) \right).
\end{align*}
Since $J(P(t))$ is convex and differentiable with respect to $P(t)$, its minimum can be found by taking derivative of $J(P(t))$ with respect to $P(t)$. Let $P'(t)$ denote the solution to $\frac{d J(P(t))}{d P(t)}=0$. It is given by
\begin{align}\label{localp}
P'(t)=\frac{-V}{\Delta t X(t)}-\frac{1}{\gamma(t)}.
\end{align}
 By  constraint \eqref{p}, to determine whether $P'(t)$ is  an optimal solution of \textbf{P3}, we consider two cases:

\subsubsection{If $X(t)<0$} In this case, $J(\cdot)$ is not a monotonic function. Define $P^*(t)$ as the optimal solution of \textbf{P3}. It is determined by comparing $P'(t)$ with the two bounds $0$ and $P_{\max}$ by constraint \eqref{p}.
In order for $P^*(t)=P'(t)$, it means $0\leq P'(t) \leq P_{\max}$.
By substituting $P'(t)$ in \eqref{localp} into \eqref{p}, the range of $X(t)$ for $P^*(t)=P'(t)$ can be found as \begin{align}\label{xcons}
\frac{-V\gamma(t)}{\Delta t} \leq X(t)\leq \frac{-V}{\Delta t}\left(\frac{1}{P_{\max}+\frac{1}{\gamma(t)}}\right)~.
\end{align}
Thus, if $X(t) < \frac{-V\gamma(t)}{\Delta t}$~, then $P^*(t)=0$. If $X(t)> \frac{-V}{\Delta t}\left(\frac{1}{P_{\max}+\frac{1}{\gamma(t)}}\right)$, then $P^*(t)=P_{\max}$.
\subsubsection{If $X(t)\geq0$} In this case, $J(\cdot)$ is a decreasing function of $P(t)$. Since $P'(t)<0$, it does not satisfy constraint \eqref{p}. Therefore, the minimum value of P3 is found by $P^*(t)=P_{\max}$~.
\endIEEEproof

\section{proof of Lemma ~\ref{lemma2}}\label{app:lemma2}
\IEEEproof
We first present the following lemma that will be used to prove  Lemma \ref{lemma2}.
\begin{lemma}\label{lemma4}
The optimal power allocation of problem {\bf P3} has the following properties:
\begin{itemize}
\item If $X(t)> 0$ then the optimal solution always chooses $P^*(t)=P_{\max}$.
\item If $X(t)<-V \zeta_{\max}$ then the optimal solution always chooses $P^*(t)=0$.
\end{itemize}
\end{lemma}
\IEEEproof
Based on Proposition \ref{prop0}, we know that if $X(t) < \frac{-V\gamma(t)}{\Delta t}~$, then $P^*(t)=0$. For $\zeta_{\max}=\frac{\gamma_{\max}}{\Delta t}$, the sufficient condition for $P^*(t)=0$ is $X(t)<-V \zeta_{\max}$. Similarly, we can derive the sufficient condition for $P^*(t)=P_{\max}$~, which is $X(t)>0$.
\endIEEEproof
Using Lemma \ref{lemma4} and Algorithm \ref{algm1}, we now prove the bounds in \eqref{xbound}. Note that by Lemma~\ref{lemma4}, when $X(t)<-V\zeta_{\max}$, in the next time slot, $X(t+1)$ in \eqref{dx} is always increasing, \ie $X(t+1)\ge X(t)$. When $-V\zeta_{\max}\leq X(t)$, by Algorithm \ref {algm1}, we have $P^*(t) > 0$. From  \eqref{dx}, the maximum possible decrease of $X(t)$ to $X(t+1)$ is when $P^*(t)=P_{\max}$ and $E_s(t)=0$, \ie using maximum transmit power and no  energy harvested. In this case, we have
\begin{align*}
X(t+1)\ge X(t)-\Delta t P_{\max}\ge -V\zeta_{\max}-\Delta t P_{\max}.
\end{align*}
Since the above inequality holds for any $t$, we conclude that $X(t)\ge X_\text{low}$ where
\begin{align} \label{xlow}
X_{\text{low}}=-V\zeta_{\max}-\Delta t P_{\max}.
\end{align}

From \eqref{sh}, we have $E_s(t)\le \min\{E_{\max}-E_b(t)+\Delta tP(t),E_{c,\max}\}$. Combining this with \eqref{dx}, we have
\begin{align}\label{lemma2:x_t+1}
X(t+1)&\le X(t)-\Delta t P(t) \nn\\
&\hspace*{-.4em} +\min\{E_{\max}-X(t)-A+\Delta tP(t),E_{c,\max}\}
\end{align}
If $X(t)>0$, by Lemma~\ref{lemma4},  $P^*(t)=P_{\max}$. This means $X(t+1)\le X(t)-\Delta tP_{\max}+E_{c,\max}\le X(t)$. Thus, $X(t+1)$ is decreasing. If $X(t)\le 0$, we have $P^*(t)\in [0,P_{\max})$. In this case, the maximum increase from $X(t)$ to $X(t+1)$ is when $P^*(t)=0$ and $E_s(t)=E_{c,\max}$. In this case, we have $X(t+1)\le X(t)+E_{c,\max}\le E_{c,\max}, \forall t$. It follows that $X(t)$ is upper bounded as $X(t)\le E_{c,\max} \triangleq X_\text{up}$.
\endIEEEproof

\section{Proof of Proposition ~\ref{prop1}}\label{app:prop1}
\IEEEproof
In order for $P^*(t)$ to be a feasible solution to {\bf P1},  $E_{b}(t)$ needs to meet the battery capacity constraint \eqref{batst}. Since $X(t)\ge X_{\text{low}}$, by \eqref{vx} and \eqref{xlow} , we have
$E_b(t)-A\ge -V\zeta_{\max} -\Delta t P_{\max}$. This means
$A\le E_b(t)+V\zeta_{\max}+\Delta t P_{\max}, \forall t$.
It follows that set
\begin{align}\label{app:A}
A= E_{\min}+V\zeta_{\max} +\Delta t P_{\max}
\end{align}
would satisfy the above constraint.
In order for $P^*(t)$ to be feasible, it requires $X(t)=E_b(t)-A\le E_{\max}-A$. Since $X(t)\le X_{\text{up}}=E_{c,\max}$, the feasibility is guaranteed if $E_{c,\max}\le E_{\max}-A$. Replacing $A$ in this inequality by the expression in \eqref{app:A}, we have
\begin{align}
V\leq \frac{E_{\max}-E_{\min}-E_{c,\max}-\Delta t P_{\max}}{\zeta_{\max}}.
\end{align}
\endIEEEproof

\section{Proof of Theorem \ref{thrm1}}\label{app:thrm1}
\IEEEproof
We adopt the approach in Lyapunov optimization theory  \cite{Neely10} to derive the bound. We first  show that there exists a stationary, randomized power control policy $\{P^r(t)\}$ for {\bf P2}, where $P^r(t)$  only depends on the current system state $\sbf(t)$, and we can bound the expected values of the cost objective  and the constraints per slot.  Using these bounds and the upper bound of drift-plus-cost metric in \eqref{ly5f}, we derive the bound in \eqref{alg_ub1}.

The following lemma can be obtained  straightforwardly  from the results in \cite{Neely10} .
\begin{lemma}\label{lemma5}
 For  system state $\sbf(t)$  i.i.d. over time, there exists a stationary randomized power control solution $P^r(t)$ that only depends on the current state $\sbf(t)$\ and  guarantees
\begin{align}
&\mathbb{E}_\Ac[R^{r}(t)]\triangleq\bar R^{r}(\Ac) = \bar R^\textrm{o}(\Ac),\label{lemma5_1} \\
&\mathbb{E}_\Ac[E_s^{r}(t)]=\mathbb{E}_\Ac[\Delta tP^{r}(t))] \label{lemma5_2}
\end{align}
where $R^{r}(t)$ and $E_s^{r}(t)$  are instantaneous rate and harvested energy under the stationary randomized solution,  $\mathbb{E}_{{\cal A}}[\cdot]$ is taken with respect to the random system state  $\sbf(t)$ conditioned on $\gamma(t)\le \gamma_{\max}$  and the randomized power  solution $P^r(t)$, and  $\bar R^{r}(\Ac)$ and  $\bar{R}^{o}(\Ac)$ are the objectives of {\bf P2} achieved under $P^r(t)$ and  under the optimal solution, respectively.
\end{lemma}

Our proposed algorithm is to solve per slot optimization problem {\bf P3}, which minimizes the upper bound in \eqref{ly5f} over all possible power control solutions, including  the optimal stationary randomized solution $P^r(t)$\ in Lemma \ref{lemma5}.
Plugging $P^r(t)$ into the right hand side of \eqref{ly5f} and by Lemma~\ref{lemma5}, we have
\begin{align}\label{ly6f}
&\Delta(X(t))-V\mathbb{E}_\Ac[R^{s}(t)|X(t)]\nonumber\\
&\leq B+X(t)\mathbb{E}_{{\Ac}}[E_s^{r}(t)-\Delta tP^{r}(t)|X(t)]-V\mathbb{E}_\Ac[R^{r}(t)|X(t)]\nonumber\\
&= B+X(t)\mathbb{E}_{{\Ac}}[E_s^{r}(t)-\Delta tP^{r}(t)]-V\mathbb{E}_\Ac[R^{r}(t)]\nonumber\\
&= B-V\bar{R}^{o}(\Ac) \nn\\
&\leq B-V\bar{R}^{\textrm{opt}}(\Ac)
\end{align}
where the first equality is due to $P^r(t)$ only depending on $\sbf(t)$, the second equality is by \eqref{lemma5_1} and \eqref{lemma5_2} of \ Lemma~\ref{lemma5}, and the last inequality is because  {\bf P2} is a relaxed version of {\bf P1} and therefore we have $ \bar{R}^{\textrm{opt}}\le \bar{R}^{o}(\Ac)$.

By  the definition of  $\Delta(X(t))$ in \eqref{lypdrift}, taking expectations of both sides in \eqref{ly6f} over $X(t)$, and summing  over  $t$ from $0$ to $T-1$,  we have
\begin{align*}
&V\sum _{t=0}^{T-1}\mathbb{E}_\Ac[R^s(t)] \nn\\
&\ge T V\bar{R}^{\textrm{opt}}(\Ac)- TB + \mathbb{E}_\Ac[L(X(T))]-\mathbb{E_\Ac}[L(X(0))] \nonumber\\
& \ge T V\bar{R}^{\textrm{opt}}(\Ac)- TB -\mathbb{E_\Ac}[L(X(0))]
\end{align*}
where the last inequality is due to $L(X(T))$ being non-negative by definition.
Dividing both sides by $VT$ and taking limits over  $T$, and noting that $L(X(0))$ is bounded, we have
\begin{align}\label{bt6f}
\lim_{T \rightarrow \infty}\frac{1}{T} \sum _{t=0} ^ {T-1}\mathbb{E_\Ac}[R^{s}(t)]&\geq \bar R^{\textrm{opt}}(\Ac)- \frac{B}{V}
\end{align}
where the left hand side of \eqref{bt6f} is $\bar R^s(V,\Ac)$.
\endIEEEproof

\section{Proof of Lemma~\ref{lemma3}}\label{app:lemma3}
\IEEEproof
To find an upper bound, we know that in each time slot, an optimum rate $R^{\topt}(t)$ is less or equal to the maximum achievable rate which is $R_{\max}(t)$:
\begin{align}
R^{\topt}(t)&\leq \log\left(1+P_{\max}\gamma(t)\right)\triangleq R_{\max}(t)\label{p80}
\end{align}
So \eqref{p80} can be written as
\begin{align}
R^{ \topt}(t)-R(t)&\leq R_{\max}(t)-R(t)\label{p81}
\end{align}
where $R(t)$ is the instantaneous rate under random event $\gamma(t) \in {\cal A}^c$.
RHS of \eqref{p81} can be written as  
\begin{align}
R_{\max}-R(t)=\log\left(\frac{1+P_{\max}\gamma(t)}{1+P^s(t)\gamma(t)}\right) \label{ub23}
\end{align}
By taking  expectation of \eqref{p81} over $\gamma(t)$ and considering \eqref{ub23}, \eqref{p81} can be written as
 \begin{align}
\mathbb{E}_{{\cal A}^c}[R^{\topt}(t)]-\mathbb{E}_{{\cal A}^c}[R(t)]\leq \mathbb{E}_{{\cal A}^c}\left[\log\left(\frac{1+P_{\max}\gamma(t)}{1+P^s(t)\gamma(t)}\right)\right]\label{p82}
\end{align}
 where $\mathbb{E}_{{\cal A}^c}[\cdot]\triangleq E[\cdot | \gamma(t) \in \Ac^c]$.
Define $g(t)\triangleq\ \mathbb{E}_{{\cal A}^c}\left[\log\left(\frac{1+P_{\max}\gamma(t)}{1+P^s(t)\gamma(t)}\right)\right]$.
Summing both sides of \eqref{p82} over T, and let $ T\to \infty$, we have
\begin{align}\label{ub27}
\lim_{T \rightarrow \infty}\frac{1}{T} \sum _{t=0} ^ {T-1}\mathbb{E}_{{\cal A}^c}[R(t)]&\geq \bar R^{\topt}- g(t)
\end{align}
 where LHS of \eqref{ub27} is $\bar R^s(\Ac^c)$. From the definition of $g(t)$, we have
\begin{align}\label{p70}
g(t)&=\int_{\gamma_{\max}}^\infty \log\left(\frac{1+P_{\max}\gamma}{1+P_{{\cal A}^c}(t)\gamma}\right)f(\gamma|\gamma>\gamma_{\max})d\gamma \nn \\
&\le \int_{\gamma_{\max}}^\infty \log\left(1+P_{\max}\gamma)f(\gamma|\gamma>\gamma_{\max}\right)d\gamma \nn\\
&= \frac{1}{1-F(\gamma_{\max})}\int_{\gamma_{\max}}^\infty \log\left(1+P_{\max}\gamma\right)f(\gamma)d\gamma \nn\\
& \triangleq G
 \end{align}
where for simplicity, we let $\gamma(t)=\gamma$, and $f(\gamma|\gamma>\gamma_{\max})$ denotes the conditional probability density function (pdf); also, $F(\gamma_{\max})=\text{Prob}(\gamma\le \gamma_{\max})$, \ie  the cumulative distribution function (cdf) of $\gamma$.
 Note that  $G <\infty$ since the integration in the second equality is finite. Thus, combining \eqref{ub27} and \eqref{p70}, we have
$\bar{R}^s(\Ac^c)\ge R^{\topt}- G$ as in  \eqref{alg_ub2}.
\endIEEEproof

\section{Proof of Corollary~\ref{cor1}}\label{app:cor1}
\IEEEproof
Note that for $h_n(t)$ being complex Gaussian with variance $\sigma_h^2$, for $n=1,\cdots,N$, $\gamma(t)$ has the $\chi$-square distribution with $2N$ degree of freedom. Recall that $\gamma(t)=|h(t)|^2/\sigma_N^2$. Define $\bar{\sigma}_h^2\triangleq \sigma_h^2/\sigma_N^2$. Thus, we have
 \begin{align}\label{f_gamma}
 f(\gamma)&=\frac{\gamma^{N-1}}{(N-1)!\bar{\sigma}_h^{2N}}e^{-\frac{\gamma}{\bar{\sigma}_h^2}} \nn\\
 1-F(\gamma_{\max})&={\frac{1}{(N-1)!}\hat\Gamma(N,\frac{\gamma_{\max}}{\bar{\sigma}_h^2})}.
 \end{align}
where  $\hat\Gamma(n,y)\triangleq\ \int_y^\infty x^{n-1}e^{-x}dx$. From the above, the upper bound $G$\ of $g(t)$  is given by
\begin{align*}
G &= C\int_{\frac{\gamma_{\max}}{\sigma_h^2}}^\infty \log\left(1+\bar{\sigma}_h^2P_{\max}\gamma \right)\gamma^{N-1}e^{-\gamma}d\gamma
\end{align*}
as shown in \eqref{G}, where $C\triangleq \left[\hat{\Gamma}\left(N,\frac{\gamma_{\max}}{\bar{\sigma}_h^2} \right)\right]^{-1}$.
A special case is when $N=1$. The channel has a Rayleigh fading and  $\gamma(t)$ has an exponential distribution. Therefore,
\begin{align}
\frac{f(\gamma)}{1-F(\gamma_{\max})}=\frac{1}{\bar{\sigma}_h^2 }{e^{-\frac{\gamma}{\bar{\sigma}_h^2}+\frac{\gamma_{\max}}{\bar{\sigma}_h^2}}}.
\end{align}
It follows that
\begin{align}
G&= \frac{1}{\bar{\sigma}_h^2}e^{\frac{\gamma_{\max}}{\bar{\sigma}_h^2}}\int_{\gamma_{\max}}^\infty \log\left({1+P_{\max}\gamma}\right)e^{-\frac{\gamma}{\bar{\sigma}_h^2}}d\gamma.
 \end{align}
By using integral by part, we have
\begin{align*}
&\frac{1}{\bar{\sigma}_h^2}\int_{\gamma_{\max}}^\infty \log\left({1+P_{\max}\gamma}\right)e^{-\gamma/\bar{\sigma}_h^2}d\gamma\nonumber\\
&\hspace*{-1.1em}=\log(1+P_{\max}\gamma_{\max})e^{-\frac{\gamma_{\max}}{\bar{\sigma}_h^2}}+\int_{\frac{\gamma_{\max}}{\bar{\sigma}_h^2}}^\infty \frac{\bar{\sigma}_h^2P_{\max}}{1+\bar{\sigma}_h^2P_{\max}\gamma}e^{-\gamma}d\gamma 
\end{align*}
For the second term above, we use the following result
\begin{align}
\int_{u}^\infty \frac{1}{\beta+x}e^{-x}dx=e^{\beta}\hat{\Gamma}(0,u+\beta)
\end{align}
 Thus, we have $G$ as in \eqref{G:N=1}
\endIEEEproof

\section{Proof of Theorem~\ref{thrm2} }\label{app:thrm2}
\IEEEproof
The achieved long-term time-averaged expected rate under Algorithm \ref{algm2} can be written as
\begin{align}\label{bound1}
\bar R^s(V,\eta) =(1-\eta)\bar R^s(V,{\cal A})+\eta {\bar R^s(\Ac^c)}
\end{align}
where $\bar R^s(V,\eta)=\lim_{T \rightarrow \infty}\frac{1}{T} \sum _{t=0} ^ {T-1}\mathbb{E}[R(t)]$.
Also, the optimal solution of  {\bf P1}  can be written as
\begin{align}\label{bound2}
\bar R^{\topt} =(1-\eta)\bar R^{\topt}(\Ac)+\eta {\bar R^{\topt}(\Ac^c)}
\end{align}
By subtracting \eqref{bound1} from \eqref{bound2}, we have
\begin{align}\label{bound3}
\bar R^{\topt}-\bar R^s(V,\eta) &=(1-\eta)~\left(\bar R^{\topt}(\Ac)-\bar R^s(V,{\cal A})\right) \nn\\
&\quad +\eta ~\left({\bar R^{\topt}(\Ac^c)}-\bar R^s(\Ac^c)\right).
\end{align}
Combining the results of  Theorem \ref{thrm1} and Lemma~\ref{lemma3}, the performance gap of Algorithm \ref{algm2} to the optimal solution for {\bf P1} in \eqref{alg_ub} follows.
\endIEEEproof

\begin{IEEEbiography}[{\includegraphics[width=1in,height=1.25in,clip,keepaspectratio]{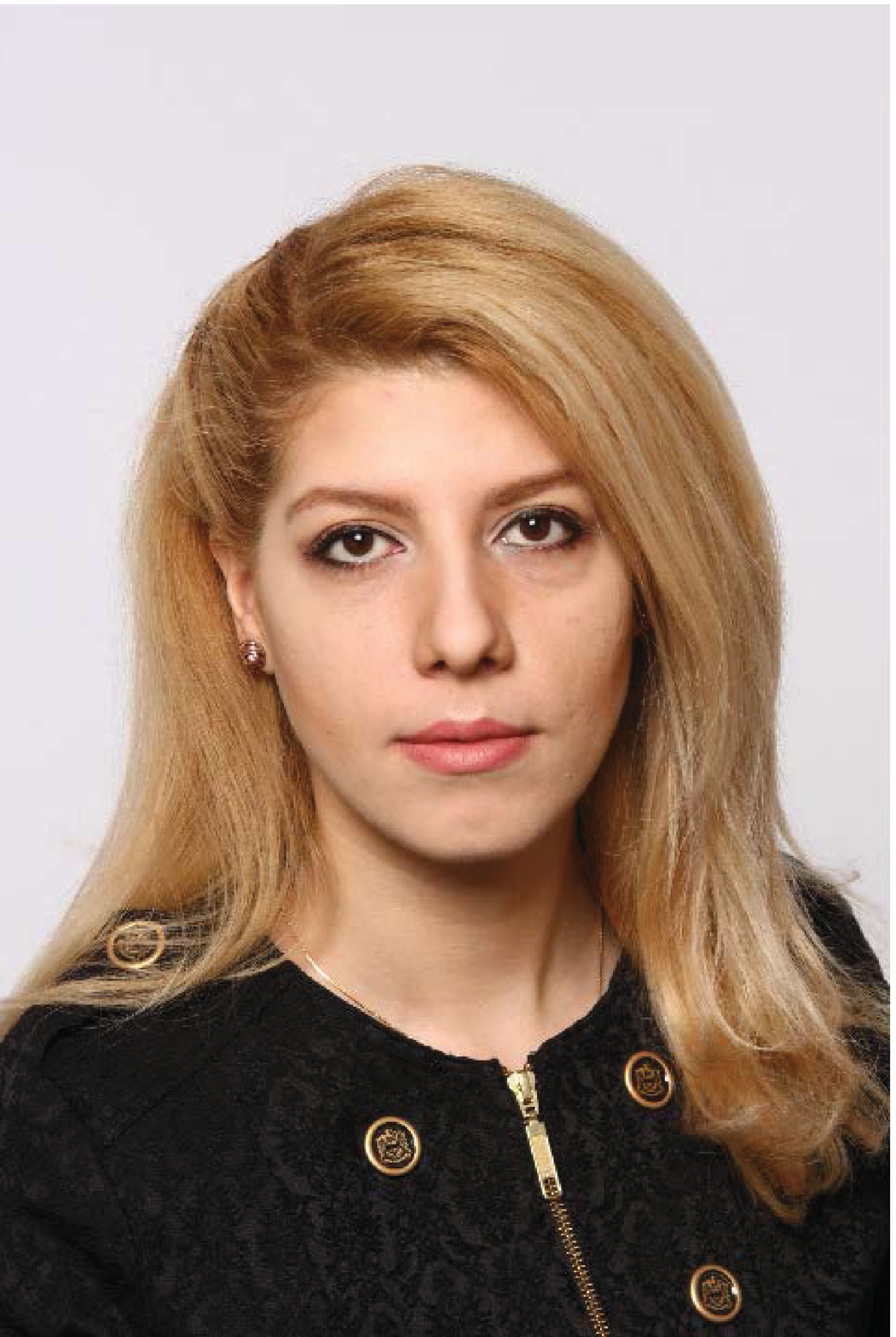}}]{Fatemeh Amirnavaei}
(S'14) received the B.Sc. and M.Sc. degrees in electrical engineering from Shariaty University and  Shahed University, Tehran, Iran, in 2006 and 2011, respectively. She is currently pursuing the Ph.D. degree in electrical engineering at University of Ontario Institute of Technology, Ontario, Canada. Her research interests include statistical signal processing in wireless communication, cooperative networks, green communication, and stochastic network optimization.
\end{IEEEbiography}
\begin{IEEEbiography}[{\includegraphics[width=1in,height=1.25in,clip,keepaspectratio]{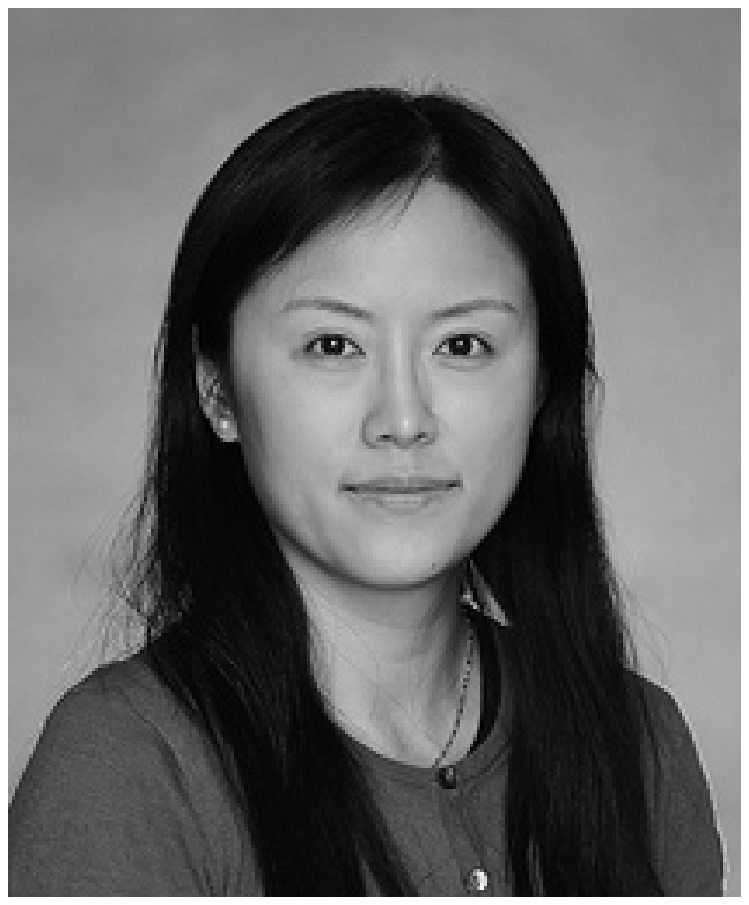}}]{Min Dong}
(S'00-M'05-SM'09) received the B.Eng. degree from Tsinghua University, Beijing, China, in 1998, and the Ph.D. degree in electrical and computer engineering with minor in applied mathematics from Cornell University, Ithaca, NY, in 2004. From 2004 to 2008, she was with Corporate Research and Development, Qualcomm Inc., San Diego, CA. In 2008, she joined the Department of Electrical Computer and Software Engineering at University of Ontario Institute of Technology, Ontario, Canada, where she is currently an Associate Professor. She also holds a status-only Associate Professor appointment with the Department of Electrical and Computer Engineering, University of Toronto since 2009. Her research interests are in the areas of statistical signal processing for communication networks, cooperative communications and networking techniques, and stochastic network optimization in dynamic networks and systems.

Dr. Dong received the Early Researcher Award from Ontario Ministry of Research and Innovation in 2012, the Best Paper Award at IEEE ICCC in 2012, and the 2004 IEEE Signal Processing Society Best Paper Award. She has served as an Associate Editor for the IEEE TRANSACTIONS ON SIGNAL PROCESSING from 2010 to 2014, and as an Associate Editor for the IEEE SIGNAL PROCESSING LETTERS from 2009 to 2013.  She has been an elected member of IEEE Signal Processing Society Signal Processing for Communications and Networking (SP-COM) Technical Committee since 2013.
\end{IEEEbiography}

\end{document}